\documentclass[a4paper, 10pt]{article}
\pdfoutput=1

\usepackage{amsmath, amssymb, bm}
\usepackage{hyperref}
\usepackage[]{graphicx}
\usepackage[caption=false]{subfig}
\usepackage[top=1in, bottom=1.2in, left=0.7in, right=0.7in]{geometry}
\usepackage{authblk}

\newcommand{\bs}{\boldsymbol}

\numberwithin{equation}{section}
\begin{document}

\title{\bf{\Large{Magneto-optic effects of the Cosmic Microwave Background}}}
\author{Damian Ejlli}

\affil{\emph{\normalsize{Department of Physics, Novosibirsk State University, Novosibirsk 630090, Russia and}}\\ \emph{\normalsize{Theory group, Laboratori Nazionali del Gran Sasso, 67100 Assergi, L'Aquila Italy}}}

\date{}

\maketitle

\begin{abstract}
Generation of magneto-optic effects by the interaction of the CMB with cosmic magnetic fields is studied. Effects which generate polarization such as the Cotton-Mouton effect, vacuum polarization and photon-pseudoscalar mixing in external magnetic field are studied. Considering the CMB linearly polarized at decoupling time, it is shown that photon-pseudoscalar mixing in external magnetic field, the Cotton-Mouton effect in plasma and the vacuum polarization in cosmic magnetic field, would generate elliptic polarization of the CMB depending on the photon frequency and magnetic field strength. Among standard magneto-optic effects, the Cotton-Mouton effect in plasma turns out to be the dominant effect in the generation of CMB elliptic polarization in the low frequency part $\nu_0 \sim 10^8-10^9$ Hz with degree of circular polarization $P_C(T_0)\simeq 10^{-10}-10^{-6}$ for magnetic field amplitude $B_{e0}\sim 1\, \text{nG}-100$ nG. The vacuum polarization in magnetic field is the dominant process in the high frequency part $\nu_0\geq 10^{10}$ Hz where the degree of circular polarization at present is $P_C(T_0)\lesssim 10^{-11}$ in the best scenario. The effect of pseudoscalar particles on the CMB polarization is also studied. It is shown that photon-pseudoscalar particle mixing in cosmic magnetic field generates elliptic polarization of the CMB as well and even in the case of initially unpolarized CMB. New limits/constraints on the pseudoscalar parameter space are found. By using current limit on the degree of circular polarization of the CMB, the upper limit of $|g_{\phi\gamma}|<4.29\times 10^{-19}(\text{G}/B_{e0})$ GeV$^{-1}$ for $m_\phi<1.6\times 10^{-14}$ eV in the weak mixing case is found. If $|g_{\phi\gamma}| < 1.17\times 10^{-24}(\text{G}/B_{e0})$ GeV$^{-1}$, a value of the order $|g_{\phi\gamma}|\simeq 10^{-26}(\text{G}/B_{e0})$ GeV$^{-1}$ for $m_\phi\simeq 1.6\times 10^{-14}$ eV in the resonant case, from large scale temperature anisotropy is obtained. Prior decoupling CMB polarization due to pseudoscalar particles is also discussed.

\end{abstract}


\vspace{1cm}

\section{Introduction}

 The interaction of light with matter and fields has been intensively studied in the literature and first quantitative studies dates back to Galileo, Newton, Faraday and Maxwell. Among the interesting effects that such interaction represents, there is one class of phenomena which includes the interaction of light (electromagnetic wave) with external electromagnetic fields. These phenomena manifest when an electromagnetic wave propagates through an external electromagnetic field that has been altered by the presence of the incident electromagnetic wave. If there is present only an external electric field, the effects that manifest are called electro-optic effects. Instead, if there is present only an external magnetic field, the effects that manifest belong to the category of the magneto-optic effects. In this work I study only the last effects.
 
Magneto-optic effects not only are important to the established physics but also allow to investigate new effects that have not been found yet. They are generally divided in three main categories that are related to transmission, reflection and absorption of the incident light by the magnetized medium. Depending on the initial polarization of the electromagnetic wave, there are essentially four magnetic-optic effects which belong to the transmission (and not only) category, the Cotton-Mouton (CM) effect, the Faraday effect and two more exotic effects which are the vacuum polarization and the mixing of photons with pseudoscalar (and also scalar) particles in external magnetic field. The reflection category includes essentially only the Kerr effect while the absorption category includes the so called molecular circular dichroism in gases and as will be shown in this work also the photon-pseudoscalar mixing in magnetic field.

In the transmission category, the CM effect has been extensively studied in the literature. It has been experimented mostly in gases, liquids, solids and to some degree even in plasma. The CM effect manifest when light propagates in a magnetized medium where the external magnetic field has a transversal component with respect to the direction of light propagation \cite{Born65}. This effect also shares a property with more exotic phenomena such as vacuum polarization (or simply QED effect) and photon-pseudoscalar mixing in magnetic field. All three effects manifest only where there is a transversal component of the external magnetic field with respect to the direction of light propagation. 

The vacuum polarization has been first proposed and studied in Ref. \cite{Heisenberg:1935qt} and since then has received much attention from both theory and experimental physics. This effect would manifest as a phase shift between the two photon states perpendicular and parallel to the external (transverse) magnetic field that eventually give rise to a birefringence effect which intensity depends on the incident electromagnetic wave frequency. One of the most important achievement from the experimental side, is to measure the acquired QED ellipticity angle of the incident light propagating through the magnetic field. Indeed, this has been the quest for the PVLAS \cite{Bakalov:1994} experiment, BFRT \cite{Cameron:1993mr} experiment and for new generation of experiments \cite{Heinzl:2006xc}. After a first claim of detection of vacuum birefringence by PVLAS experiment \cite{Zavattini:2005tm}, there is still a long way to achieve the required apparatus sensitivity in order to measure the QED predicted ellipticity which is by more than an order of magnitude smaller than the current apparatus sensitivity. At current status, apparatus sensitivity is contaminated with not well understood background noise, must probably from the same apparatus and new methods have also been proposed \cite{Zavattini:2012zs}.

The birefringence effect predicted by QED can also be mimicked by another magneto-optical effect, namely the photon-pseudoscalar/scalar mixing in magnetic field. In fact, as it will be shown in this work, mixing of photons with pseudoscalar particles gives rise to both birefringence and dichroism effects. Therefore experiment such as PVLAS,  BFRT etc., can in principle find pseudoscalar particles such as axions, ALPs, scalar bosons if the induced birefringence or dichroism signal is bigger than the QED expected signal. Other important experiments that aim to find exotic pseudoscalar particles include the CAST and IAXO experiments \cite{Dafni:2006pc}, ADMX experiment \cite{Asztalos:2009yp} and ALPS-II \cite{Bahre:2013ywa}.

Among all magneto-optic effects, the Faraday effect has received much attention in astronomy and cosmology. It manifest when an initial linearly polarized electromagnetic wave interacts with an external magnetic field that has a longitudinal component along the wave propagation direction. This coupling makes possible the rotation of the polarization plane of the incident electromagnetic wave and the rotation angle is proportional to $B_e d$ where $B_e$ is the strength of the external magnetic field and $d$ is the length of the path. Consequently, the Faraday effect has been widely used in radio astronomy as a probe of cosmic magnetic fields, in galaxy clusters and also in the intergalactic space \cite{Kim:1991zzc}. Measurements of the rotation angle of light received from galaxy clusters confirm the presence of a magnetic field inside them, with a magnitude of about few $\mu$G. In the intergalactic space, present studies would suggest a weaker large scale magnetic field with upper limit magnitude $B_e\lesssim 3-1380$ nG, see for example current limits by Planck collaboration \cite{Ade:2015cva} where limits of the order of 1380 nG are set from Faraday effect. On the other hand non observation of gamma rays emission from intergalactic medium due to injection of high energy particles by blazars would suggest a lower value on the strength of extragalactic magnetic field  $B_e\geq 10^{-16}-10^{-15}$ G \cite{Neronov:1900zz}. The origin of this field is still unknown and present studies suggest that it may have been created during structure formation or it may have a primordial origin,  see Ref. \cite{Grasso:2000wj} for a review on cosmic magnetic fields structure and Ref. \cite{Ade:2015cva} for current updated limits/constraints by the Planck collaboration. In this work it is assumed that the magnetic field has a primordial origin and its amplitude is a slowly varying function of space-time coordinates, namely a slowly varying inhomogeneous field in space and time which can be also stochastic in nature, see sec. \ref{sec:7} for details.

In connection with the CMB physics, the Faraday effect has been used to probe the existence of primordial magnetic field \cite{Kosowsky:1996yc} present at the decoupling time since it would rotate the polarization plane of the CMB. In fact, it is well known by now that the CMB posses a very small linear polarization that is believed to have been generated at the decoupling time due to Thomson scattering of CMB photons on electrons. Such a polarization is generated because of temperature anisotropies present at the decoupling epoch that eventually generate a position dependent photon intensity on the surface of the last scattering \cite{Bond:1987ub}. Consequently, Thomson scattering of an anisotropic background of photons on electrons would generate linear polarization of the CMB with non zero Stokes parameters $Q$ and $U$ \cite{Kosowsky:1994cy}, \cite{Hu:1997hv}. 

In general, the linear polarization pattern of the CMB can be decomposed in two modes with opposite parity, the so called E-modes (or gradient modes G) which are the dominant component of the linear polarization and B-modes (or curl modes C) which are the subdominant component of linear polarization, see Refs. \cite{Kamionkowski:1996ks}.  Scalar density fluctuations of the cosmological plasma during the decoupling time generate E-modes only which are consistent with observations, while vector and tensor perturbations can generate both E-modes and B-modes. The generation of B-modes is induced by tensor perturbations (gravitational waves) \cite{Crittenden:1993wm}, Faraday rotation of the CMB \cite{Kosowsky:1996yc}, gravitational lensing of the E-mode component \cite{Zaldarriaga:1998ar} and due to primordial magnetic fields \cite{Mack:2001gc} via perturbations sourced by the magnetic field. In general, the spectrum of B-modes is described in multipole moments $l$ of spherical harmonics used to describe linear polarization. The location of the peak signal of B-modes as function of $l$ would give the possibility to distinguish between signals generated by different sources of B-modes. 

So far, much of attention on the CMB polarization has been focused mostly on the linear polarization. This fact, mostly has been influenced by the first experimental observation of E-modes (due to primordial adiabatic scalar fluctuations) by DASI, WMAP and BOOMERANG collaborations \cite{Kovac:2002fg} and also partially by the fact that many inflationary models predict an almost scale invariant spectrum of gravitational waves, which as already mentioned above, can produce B-modes which are believed to be the `holy grail' of the inflationary theory. Moreover, since Thomson scattering is the most frequent type of scattering in the early universe and because it generates only linear polarization, other types of CMB polarization have been to some extent obscured and the $V$ Stokes parameter has become essentially the `lost along the way' parameter. However, it is well known that light can have two additional types of polarization, circular and elliptic which translate into a nonzero Stokes parameter $V$.

After this premise on the CMB linear polarization, several questions come spontaneously. Does the CMB posses only linear polarization? Does it have any degree of circular polarization? If yes, what are the generating mechanisms? Even though, there is not urgency on the study of CMB circular polarization, since the discovery of the CMB, there have been several attempts in the past and also at the present to experimentally measure it. Moreover, since CMB linear polarization has already been detected, the next step would be that of the study of circular polarization which as I will show in this paper is generated by very interesting mechanisms which are extremely important to the fundamental physics.

The first studies on the CMB circular polarization were done in connection with studies on anisotropic expansion of the universe which are characterized by some type of Bianchi models \cite{negroponte:80}. Other studies on generation of CMB circular polarization include; interaction of the CMB with a vector field via a Chern-Simons term\cite{Alexander:2008fp}, non commutative geometry \cite{Bavarsad:2009hm}, electron-positron scattering in magnetized plasma at decoupling time \cite{Giovannini:2010ar}, propagation of CMB photons in magnetic field of supernova remnants of the first stars \cite{De:2014qza}, photon-pseudoscalar mixing in magnetic field domains \cite{Agarwal:2008ac}, scattering of the CMB photons with cosmic neutrino background \cite{Mohammadi:2013dea}. For a recent review on other CMB circular polarization mechanisms see Ref. \cite{King:2016exc}. The first experimental attempts to measure the circular polarization of the CMB were done in Ref. \cite{smooth:83} where no evidence for CMB circular polarization was found and only constraint on the degree of circular polarization was set. The current upper limit on the CMB circular polarization has been set by the MIPOL experiment \cite{Mainini:2013mja}, $P_C(T_0)\lesssim 7\times 10^{-5}-5\times 10^{-4}$ at the frequency 33 GHz and at angular scales between $8^\circ$ and $24^\circ$.

In this work I study the impact of magneto-optic effects on the CMB polarization in the presence of cosmic magnetic fields where I mostly concentrate on generation of CMB circular polarization. A systematical study of the most important magneto-optic effects in the generation of a net CMB elliptic (circular and linear) polarization is done. By including all magneto-optic effects mentioned above, I derive the equations of motion for the Stoke's parameters which form a coupled system of differential equations. I use a density matrix approach to study the mixing of different magneto-optic effects and then solve the equations of motion by using perturbation theory. It turns out that among CM and vacuum polarization effects, the CM effect in plasma is the most promising effect in generation of elliptic polarization in the low frequency part of the CMB, while in the high frequency part, the vacuum polarization is the dominant one. I also will use current limit on the degree of circular polarization, to set new limits on the mass and coupling constant of pseudoscalar particles. In connection with CMB circular polarization, I calculate its magnitude in terms of degree of circular polarization at present $P_C(T_0)$ and compare with experimental result(s). Generation and evolution of CMB E-mode and B-mode generated by the above mentioned effects is not studied in this work.

This paper is organized as follows: In Sec. \ref{sec:2}, I derive the equations of motion for the photon and pseudoscalar fields in an expanding universe and introduce the photon polarization tensor in magnetized medium which describes forward scattering of photons. In Sec. \ref{sec:3}, I study the equations of motion for the density matrix in the case of open systems and establish the connection between the system Hamiltonian and the field mixing matrix. In Sec. \ref{sec:4}, I find the equations of motion for the density matrix in an expanding universe and solve them in the case of vacuum polarization and CM effects. In Sec. \ref{sec:5}, I present the equations of motion for the density matrix in the case when the contribution of the pseudoscalar field is included and introduce the concept of generalized Stokes parameters. Then I find perturbative solutions of the reduced Stokes vectors in transverse magnetic field. In Sec. \ref{sec:6}, I study the generation of CMB circular polarization in the case of photon-pseudoscalar particle mixing in transverse magnetic field and set new limits on the pseudoscalar parameter space. In Sec. \ref{sec:7}, I conclude. In this work I use the metric with signature $\eta_{\mu\nu}=\textrm{diag}(1, -1, -1, -1)$ and work with the natural (rationalized) Lorentz-Heaviside units ($k_B=\hbar=c=\varepsilon_0=\mu_0=1$) with $e^2=4\pi \alpha$.

\section{Equations of motions in an expanding universe}
\label{sec:2}

In this section we derive the equations of motion for the photon and pseudoscalar fields propagating in a magnetized  medium in the framework of the Friedemann-Robertson-Walker (FRW) metric. To start with, we write the effective action of the photon and pseudoscalar fields in curved spacetime
\begin{align}
\mathcal S_\text{eff} &=\int d^4x\sqrt{-g}\left(-\frac{1}{4} F_{\mu\nu}F^{\mu\nu}-\frac{1}{2}\int d^4x^\prime\,A_\mu (x)\Pi^{\mu\nu}(x, x^{\prime})A_\nu(x^{\prime})+\frac{1}{2}\partial_\mu\phi\, \partial^\mu \phi \right.\nonumber \\ & \left. -\frac{1}{2}m_\phi^2\,\phi^2+\frac{g_{\phi\gamma}}{4}\phi\,F_{\mu\nu}\tilde F^{\mu\nu}\right),
\end{align}
where $F_{\mu\nu}$ is the total electromagnetic field tensor, $\Pi^{\mu\nu}$ is the photon polarization tensor in medium, $\phi$ is the pseudoscalar field, $m_\phi$ is the mass of the pseudoscalar field, $g$ is the metric determinant and $A^\mu$ is the photon vector potential. By varying the action with respect to the electromagnetic field $A^\nu$ and pseudoscalar field $\phi$, the equations of motion are
\begin{eqnarray}\label{eq-mo}
\Box A^\nu-\nabla_\mu(\nabla^\nu A^\mu)-\int d^4 x^\prime\, \Pi^{\mu\nu}(x, x^\prime)\,A_\mu(x^\prime) &=& g_{\phi\gamma} (\partial_\mu \phi) \tilde F^{\mu\nu},\nonumber\\
(\Box+m_\phi^2) \phi=\frac{g_{\phi\gamma}}{4}F_{\mu\nu}\tilde F^{\mu\nu},
\end{eqnarray}
where $\nabla_\mu \tilde F^{\mu\nu}=0$, $\Box=\nabla_\mu \nabla^\mu$ is the d'Alambertian operator in curved space, $x^\mu=(t, \bs x)$, $\nabla_\mu$ is the covariant derivative and $\nabla_\mu\phi=\partial_\mu\phi$. In this work we consider the case of flat ($\kappa=0$) FRW metric with line element $ds^2=dt^2-d\bs x^2(t)$,
where $t$ is the cosmological time and $\bs x$ is the physical spatial coordinate. The only non zero components of the affine connection in the FRW metric are $\Gamma_{0j}^i=(\dot a/a)\delta_{ij}$ and $\Gamma_{ij}^0=\dot a a\delta_{ij}$ where $a(t)$ is the cosmological scale factor.

In general the electromagnetic field tensor $F_{\mu\nu}$ is given by the sum of the field tensor of  incident photon field and of field tensor corresponding to the external magnetic field. In most cases the electromagnetic field tensor corresponding to the external magnetic field is the dominant term. Considering the photon propagation in an external magnetic field, the equations of motion \eqref{eq-mo} for the components of vector potential $\bs A^i$ and pseudoscalar field $\phi$ in the Coulomb gauge and in the unperturbed FRW metric\footnote{In the Coulomb gauge there is also the equation of motion for $A^0$ (scalar potential) which is proportional to $(\nabla\cdot \phi)\bs B_e$ for a globally neutral medium. In this case the mixing problem has four coupled differential equations in the case when $\bs B_e$ is not transverse. However, the effect of this equation to the mixing problem is very small, namely of the order $(g_{\phi\gamma} B_{eL})^2$ where $B_{eL}$ is the magnitude of the longitudinal component of $\bs B_{e}$ and therefore we can safely neglect this equation for our purposes \cite{Das:2004ee}.} are
\begin{eqnarray}\label{eq-A-ph}
(\partial_t^2-\nabla^2+3H\,\partial_t)\bs A^i+\int d^4 x^\prime\,\Pi^{ij}(x, x^\prime) \bs A_j(x^\prime) &=&-g_{\phi\gamma}(\partial_t\phi) \bs B_e^i,\nonumber\\
(\partial_t^2-\nabla^2+3H\,\partial_t+m_\phi^2)\phi &=& g_{\phi\gamma}\partial_t\bs A_i\cdot\bs B_e^i.
\end{eqnarray}
We may notice that there is an extra term in the equations of motion \eqref{eq-A-ph} with respect to the Minkowski flat space-time for the photon and pseudoscalar fields, that is $3 H\partial_t$ where $H=\dot a/a$ is the Hubble parameter. This term is the so called Hubble friction that is responsible for the damping of the fields in an expanding universe.

We look for single wave vector solutions of Eqs. \ref{eq-A-ph} of the form
\begin{equation}\label{expansion}
\bs{A}_j(\bs x, t)=\sum_\lambda\,A_\lambda (\bs k, t) \bs e_j^\lambda(\hat{\bs n})\, e^{i\bs{k\cdot x}},\quad \phi (\bs x, t)=\phi(\bs k, t) e^{i\bs{k\cdot x}},
\end{equation}
where $\bs k$ is the photon wave vector, $\bs e^\lambda$ is the photon polarization vector and $\lambda$ is the photon polarization index. For simplicity, we consider an electromagnetic wave propagating along the observer's $\hat {\bs z}$ axis with $\bs k=(0, 0, k)$, $k=|\bs k|$. Without any loss of generality we choose the external magnetic field in the $xz$ plane with coordinates $\bs B_e=(B_e\sin(\Phi), 0, B_e\cos(\Phi))$ where $\Phi$ is the angle between the magnetic field direction and photon wave vector $\bs k$, $\cos(\Phi)=\hat{\bs B_e}\cdot \hat{\bs n}$ with $\hat{\bs n}=\bs k/k$. Given the symmetry of the problem, only the transverse part of the external magnetic induces photon-pseudoscalar mixing. Inserting the expansion \eqref{expansion} into the equations of motion \eqref{eq-A-ph} we obtain
\begin{eqnarray}\label{initial-sys}
\left(i\partial_t-k+\frac{3}{2}iH\right) A_+ (k, t)+ M_+(k)A_+(k, t)+iM_F(k)A_\times(k, t) &=& 0,\nonumber\\
\left(i\partial_t-k+\frac{3}{2}iH\right)A_\times (k, t)+M_\times(k) A_\times(k, t)-iM_F(k)A_+(k, t)+i M_{\phi\gamma}(k)\phi(k, t)  &=& 0,\\
\left(i\partial_t-k+\frac{3}{2}iH\right) \phi (k, t)- iM_{\phi\gamma}(k)A_\times(k, t) +M_\phi(k)\, \phi(k, t) &=& 0,\nonumber
\end{eqnarray}
where we used the WKB approximation, namely that $\partial_t |A_\lambda|\ll \omega |A_\lambda|$ and $\partial_t |\phi|\ll \omega\,|\phi|$. Indeed, this approximation is well satisfied since the variation in time of external potential (that is proportional to the external magnetic field amplitude) due to universe expansion is much smaller than photon/pseudoscalar frequency. We also used the fact that photons and pseudoscalar particles are assumed to be relativistic and expanded the operator $(\partial_t^2-\nabla )\simeq 2k(-i\partial_t+k)$. The system of Eqs. \ref{initial-sys} can be written in a matrix form as follows\footnote{Similar equations in Minkowski space-time are found in Ref. \cite{Raffelt:1987im}}
\begin{equation}\label{matrix-eq}
\left(i\partial_t-k+\frac{3}{2}iH\right) \left(\begin{matrix}
  A_+\\
  A_\times\\
   \phi\\
 \end{matrix}\right)\bs I+
\left(\begin{matrix}
   M_{+} & iM_F & 0 \\
  -iM_F  &   M_{\times} & i M_{\phi\gamma} \\
  0 & -i M_{\phi\gamma} &  M_\phi\\
   \end{matrix}\right)
\left(\begin{matrix}
  A_+\\
  A_\times\\
   \phi\\
 \end{matrix}\right)=0.
\end{equation}
Here $A_+$ is the photon state perpendicular to the transverse part of $\bs B_e$, $A_\times$ is the photon state parallel to the transverse part of $\bs B_e$ and $\bs I$ the identity matrix. The photon state labels ($+, \times$) which essentially correspond to the $y$ and $x$ components of $\bs A^i$ should not be confused with the gravitational wave polarization states ($+, \times$). The diagonal elements of the mixing matrix $M$ in Eqs. \eqref{matrix-eq} are $M_+=-\Pi^{22}/(2 k), M_\times=-\Pi^{11}/(2k)$ and $M_\phi=-m_\phi^2/(2k)$, while the off diagonal elements are $M_{\phi\gamma}=g_{\phi\gamma}B_e\sin(\Phi)/2$ and $iM_F=-\Pi^{12}/(2k)$ is the term that corresponds to the Faraday effect. The elements of the photon polarization tensor\footnote{The elements of the photon polarization tensor in magnetized, non relativistic and non degenerate electron plasma calculated in Ref. \cite{D'Olivo:2002sp}, include only the Faraday effect and CM effect in plasma. They do not include the contribution of vacuum polarization in magnetic field and CM effect in gases.} $\Pi^{11}, \Pi^{22}, \Pi^{12}$ and $\Pi^{21}$ are calculated in momentum space  \cite{D'Olivo:2002sp} where we took the adiabatic limit $t^\prime\rightarrow t$. Their expressions will be given explicitly in the next sections. As far as concerns the nature of the large scale magnetic field, in this work is assumed that it has fixed direction in the sky and its amplitude $B_e$ is a slowly varying function of space coordinates, namely homogeneous or almost homogeneous in space. On the other hand due to universe expansion, the field amplitude changes in time, namely the large scale magnetic field is non stationary.

\section{Open systems}
\label{sec:3}

In this section we consider the case when photons (for example the CMB) are considered to interact with a medium, which for example can be magnetic field and cosmological plasma. Our goal is to find the equation of motion for the density matrix which in general is not trivial. Here we are interested in quantities that are proportional to the amplitude square of the fields and because we want to study the mixing of CMB photons with pseudoscalar particles, the density matrix approach is the most adapted in this situation. Another fact in favor of this approach is that the CMB is almost unpolarized where the statistical mixture is maximal and the description of such a state demands the use of the density matrix. In the case when a system couples to another system, we are dealing with open systems that exchange energy and matter between each other. Therefore, in the case of photons interacting with plasma and magnetic field, the photon number is not in general conserved and the most important processes that can change their number, in the case that we treat in this work, is photon-pseudoscalar particle mixing in the cosmological plasma.

In the general case of an open quantum system, the equations of motion for the total density matrix, in the Schr\" odinger picture, are given by the von Neumann equation
\begin{equation}\label{L-N}
i\frac{\partial\rho}{\partial t}=[H_T, \rho],
\end{equation}
where $\rho$ is the total density matrix of the system and $H_T$ is the total Hamiltonian (not to be confused with the Hubble parameter of the next section). The total system, is in general the sum of a quantum system $S$ which is coupled to another quantum system $B$ which is called the environment or bath, namely $S+B$. The total system considered here is assumed to be closed, following Hamiltonian dynamics. The state of the system $S$, which we call the photon-pseudoscalar system, will change as a consequence of its internal dynamics and because of the interaction with its surroundings. The interaction leads to system-environment correlations, such that state changes of $S$, can no longer be represented in terms of unitary Hamiltonian dynamics. In this context, the photon-pseudoscalar system $S$ is also called a reduced system.

Suppose that $\mathcal H_S$ is the Hilbert space of the photon-pseudoscalar system $S$ and $\mathcal H_B$ is the Hilbert space of the environment.   The Hilbert space of the total system $S+B$ would be the tensor product $\mathcal H=\mathcal H_S\otimes \mathcal H_B$ and the total Hamiltonian has the general form $ H_T=H_S\otimes I_B+I_S\otimes H_B+H_I(t)$,
where $H_S$ is the free Hamiltonian of the reduced system, $H_B$ is the free Hamiltonian of the environment, $H_I$ is the interaction Hamiltonian between the two systems $S$ and $B$ and $I_B, I_S$ are identity operators in their corresponding Hilbert spaces. If we are interested in the observables of only system $S$, we can define the density operator of such system by taking the partial trace on the total density operator of the system $\rho$ as follows
\begin{equation}\label{reduced-rho}
\rho_S=\textrm{Tr}_B[\rho],
\end{equation}
where $\rho_S$ is the density operator of the system $S$ (photon-pseudoscalar system) and Tr$_B$ is the partial trace over the environment degrees of freedom. Inserting \eqref{reduced-rho} into the von-Neumann equation we get
\begin{equation}\label{dens-eq}
i\frac{\partial\rho_S}{\partial t}=\textrm{Tr}_B[H_T, \rho].
\end{equation}

Equation \eqref{dens-eq} is a general result which describes the evolution in time of the reduced system interacting with an \emph{arbitrary} medium. The explicit form of the expression Tr$_B[H_T, \rho]$ on the r. h. s. of Eq. \eqref{dens-eq}, generally depends on different processes that appears in a specific problem and on type of fields that interact with the system. In our case we deal with photons that interact with different particle fields in the cosmological plasma, such as electrons, positrons, protons, light nuclei, cosmic magnetic field and in principle with other exotic particles. We refer to these fields as background fields and the calculation of the expression Tr$_B[H_T, \rho]$ would be quite involved. In fact, as one may realize at this point, there are essentially two ways on writing down the equations of motion for $\rho_S$. The first possibility would be to start from the general expression \eqref{dens-eq} and use the Hamiltonian of the total system and calculate the commutator with $\rho$ by taking the partial trace over $B$. The second possibility would be to start with the effective action and derive the equations of motions for the fields by including the effective polarization tensor for photons and their interaction with the pseudoscalar field. In the latter case, one can derive a Schr\"odinger type equation, which dynamics is governed by an effective Hamiltonian that is given by the mixing matrix $M$ and a `damping' term due to the Hubble friction as in \eqref{matrix-eq}. Obviously, the second method is more convenient since it bypasses all the tedious procedure in calculating the r. h. s. of Eq. \eqref{dens-eq}. Similar approach has been widely used also in neutrino physics \cite{Stodolsky:1986dx} and it is still the most used approach on calculating oscillation probabilities in presence of damping. However, the second approach mentioned above is an approximation of the first method and should not be sought as the most standard procedure.

All told, we work under the approximation 
\begin{equation}\label{trace}
\textrm{Tr}_B[H_T, \rho]\approx [M, \rho_S]-i \{D, \rho_S\},
\end{equation}
where $M$ is the field mixing matrix that is already `traced out' since it includes the effect of background fields on photons, photon-pseudoscalar interaction and $D$ is a `damping' matrix that is given by $D=(3/2)H\bs I$ where $H$ is the Hubble parameter. On the right hand side of \eqref{trace} instead of the total system density matrix appears only the reduced system density matrix $\rho_S$. This is due to the fact that the coupling between $S$ and $B$ is weak such that the influence of $S$ on $B$ is very small (the so called Born approximation). In such case, at a given time $t$ one can approximate\footnote{It is import to stress that this approximation does not imply that there are no excitations in the background fields.} $\rho(t)\approx\rho_S\otimes \rho_B$ \cite{Petruccione}. Consequently, the equation of motion for the density matrix becomes
\begin{equation}\label{dens-eq-1}
\frac{\partial\rho_S}{\partial t}=-i[M, \rho_S]- \{D, \rho_S\},
\end{equation}
where the first term in \eqref{dens-eq-1} describes an unitary evolution and the second term describes the `damping' of fields in an expanding universe.

\section{Photon polarization effects}
\label{sec:4}

In this section we focus on the case when the mixing matrix 
$M$ is not stationary and look for solutions of equations of motion of the density matrix, Eq. \eqref{dens-eq-1}. Indeed, it is more convenient to work with the density matrix than the wave equation, Eq. \eqref{matrix-eq}.  In this section we consider the case of missing pseudoscalar field.  As already mentioned, in the presence of an external magnetic field, excluding for the moment the case of photon-pseudoscalar mixing, there are essentially other three magneto-optic effects which depend on the external magnetic field direction and which are proportional to its strength.

In the presence of vacuum polarization, CM and Faraday effects the equation of motion of the density matrix in terms of the Stokes parameters\footnote{For the definition of photon density matrix and its connection with the Stokes parameters, see Appendix \ref{app:1}} are given by
\begin{equation}
\dot I = -3HI,\quad 
\dot Q = -2 M_{F}U-3HQ\label{V},\quad
\dot U =2 M_F Q +(M_+-M_\times)V-3HU,\quad
\dot V= -(M_+-M_\times) U-3HV
\end{equation}
where we wrote the elements of the density matrix in terms of the Stokes parameters. In general, the r. h. s. of \eqref{V} would depend on the temperature $T$ rather than $t$. Therefore, expressing the time as $t=t(T)$, the time derivative in the FRW metric becomes $\partial_t=-HT \partial_T$ where $H=-\dot T/T$. The system \eqref{V} can be written in the following matrix form
\begin{equation}\label{stokes-eq}
S^\prime(T)=A(T)\cdot S(T)+(3/T)\bs I\cdot S(T),
\end{equation}
where $S$ is the Stokes vector\footnote{The Stokes `vector' defined here is not really a vector in the mathematical sense since its components do not transform as those of an usual vector under coordinate transformation. The letter $S$ used from now on for the Stokes vector should not be confused with the letter used to denote the photon-pseudoscalar particle system $S$ of the previous section.} defined as $S=(I, Q, U, V)^\mathrm{T}$ and $A(T)$ is a matrix defined as
\begin{equation}\nonumber
A(T)=\frac{1}{HT}\left(
\begin{matrix}
0 & 0 & 0 & 0\\
0 & 0 & 2M_F(T) & 0\\
0 & -2M_F(T) & 0 & -\Delta M(T)\\
0& 0 & \Delta M(T) & 0
\end{matrix}\right),
\end{equation}
where $\Delta M(T)\equiv M_+(T) - M_\times(T)$.

The system \eqref{stokes-eq} is a first order system of linear differential equations with variable coefficients. Even though the matrix $A(T)$ that enters \eqref{stokes-eq} looks very simple, generally the system \eqref{stokes-eq} has no closed form of solutions. However, it is possible to find analytic solutions by using the perturbation theory. Indeed, as we will see in what follows, for the parameter space of the  photon/pseudoscalar momentum $k$ and magnetic field strength $B_e$ which we study in this work, one has in most cases the condition $M_F\gg |\Delta M|$. This condition on the other hand depends on $\Phi$ and for values of $\Phi\rightarrow \pi/2$, the Faraday term vanishes. In this case the condition $M_F\gg |\Delta M|$ would not be valid anymore. Therefore, we focus on for the moment in the case when $\Phi\neq \pi/2$ in such way that condition $M_F\gg |\Delta M|$ holds and split the matrix $A(T)$ in the following way
\begin{equation}
A(T)=A_0(T)+\epsilon A_1(T)=\frac{1}{HT}\left(\begin{matrix}
0 & 0 & 0 & 0\\
0 & 0 & 2M_F(T) & 0\\
0 & -2M_F(T) & 0 & 0\\
0 & 0 & 0 & 0
\end{matrix}\right)+\epsilon\left(\begin{matrix}
0 & 0 & 0 & 0\\
0 & 0 & 0 & 0\\
0 & 0 & 0 & -\tilde G(T)\\
0 & 0 & \tilde G(T) & 0
\end{matrix}\right),
\end{equation}
where we wrote $\Delta M(T)/(HT)=\epsilon\,\tilde G(T)$ and $\epsilon\ll 1$ is a parameter that depends on momentum $k$, magnetic field strength $B_e$ and on the angle $\Phi$. Here $\tilde G(T)$ is a function that depends only on the temperature $T$. The numerical factor in the product $\epsilon\,\tilde G(T)$ is included in the parameter $\epsilon$. The expression of $\epsilon$ will be given in the next sections. The second term which appears in  \eqref{stokes-eq} corresponds to the Hubble friction and its contribution to $S(T)$ appears as a damping factor of the form $\exp[-3\int_T (1/T^\prime) dT^\prime]$ and it is common to all components of the Stokes vector. The easiest way to see it, is by observing that matrix $A(T)$ commutes with $(3/T)\bs I$ for every $T$. For the moment we concentrate on the solution of Eq. \eqref{stokes-eq} without the damping term and include it in the final result.

We look for solution of the Stokes vector up to first order in $\epsilon$ as follows
\begin{equation}\label{expansion-1}
S(T)= S_0(T)+\epsilon S_1(T)+O(\epsilon^2) + ...
\end{equation}
Inserting expansion \eqref{expansion-1} into Eq. \eqref{stokes-eq} and collecting the appropriate terms we get the following matrix equations
\begin{eqnarray}
S_0^\prime(T) &=&A_0(T)S_0(T)\label{order-0},\\
S_1^\prime(T) &= & A_0(T)S_1(T)+A_1(T)S_0(T)\label{order-1}.
\end{eqnarray}
We may observe that for different cosmological temperatures, the commutator of $[A_0(T_1), A_0(T_2)]=0$ which allows us to find the following exact solution for $S_0(T)$
\begin{equation}\nonumber
S_0(T)=\left(\begin{matrix}
\cos[F(T)] & -\sin[F(T)] & 0\\
\sin[F(T)] & \cos[F(T)] & 0\\
0 & 0 & 1
\end{matrix}\right)S_0(T_i),
\end{equation}
where $T_i$ is the initial temperature and 
\begin{equation}\label{fa-F}
F(T)\equiv 2\int_{T}^{T_i} \frac{M_F(T^\prime)}{H(T^\prime) T^\prime} dT^\prime,
\end{equation}
where $T< T_i$ is the CMB temperature after the decoupling time.
We may observe that the homogeneous part of Eq. \eqref{order-1} has the same solution as Eq. \eqref{order-0} with the replacement $S_0(T)\rightarrow S_1(T)$. The non homogeneous part of Eq. \eqref{order-1} can be solved with the method of the variations of constants. Performing several algebraic operations that involve matrix exponentiations, collecting all the appropriate terms together and including the term corresponding to the Hubble friction we get the following solutions for the components of the Stokes vector to the first order in $\epsilon$:
\begin{align}
(T_i/T)^3 I(T) &= I_i, \label{sol-I}\\
(T_i/T)^3Q(T) &=\cos[F(T)]Q_i-\sin[F(T)]U_i+\left(\cos[F(T)]\int_{T}^{T_i}\epsilon\,\tilde G(T^\prime)\sin[F(T^\prime)]dT^\prime\right.\nonumber\\ &-\left.\sin[F(T)]\int_{T}^{T_i}\epsilon\,\tilde G(T^\prime)\cos[F(T^\prime)]dT^\prime\right)V_i,\label{sol-Q}\\
(T_i/T)^3U(T) &=\sin[F(T)]Q_i+\cos[F(T)]U_i+\left(\sin[F(T)]\int_{T}^{T_i}\epsilon\,\tilde G(T^\prime)\sin[F(T^\prime)]dT^\prime\right.\nonumber\\ &+\left.\cos[F(T)]\int_{T}^{T_i} \epsilon\,\tilde G(T^\prime)\cos[F(T^\prime)]dT^\prime\right)V_i,\label{sol-U}\\
(T_i/T)^3V(T) &=-\left(\int_{T}^{T_i} \epsilon\,\tilde G(T^\prime)\sin[F(T^\prime)]dT^\prime\right)Q_i-\left(\int_{T}^{T_i} \epsilon\,\tilde G(T^\prime)\cos[F(T^\prime)]dT^\prime\right)U_i+V_i\label{sol-V},
\end{align}
where $I_i, Q_i, U_i, V_i$ are the values of the Stokes parameters at temperature $T=T_i$.

There are several interesting considerations that can be made about \eqref{sol-Q}-\eqref{sol-V}. In the first place we may notice that each solution is proportional to the initial values of the Stokes parameters $Q_i, U_i$ and $V_i$, as one would expect from a first order system of linear differential equations. This implies that if the initial conditions are all zero, as for example in the case of unpolarized light, it would remain unpolarized during the universe expansion. If this is the case, the Faraday effect, the vacuum polarization and CM effect would not have any impact on the CMB polarization at all. The only way that these effects can have an impact on the CMB polarization, would be if the CMB is initially polarized.  As already mentioned in the introduction section, Thomson scattering would generate CMB linear polarization only if there are anisotropies in the CMB temperature (or intensity). If the incident light is initially unpolarized and anisotropic, Thomson scattering generates outgoing polarized light with non zero Stokes parameters $I$ and $Q$ while $V=U=0$. This is a general property of Thomson scattering for anisotropic incident light. Since the parameters $Q$ and $U$ depends on the coordinate system, one can rotate the system to a common one, in such a way to have $U\neq 0$. It can be shown that in the rotated system, the temperature anisotropy of the CMB generates non zero initial Stokes parameters $Q_i$ and $U_i$ at the decoupling time \cite{Kosowsky:1994cy}, \cite{Kosowsky:1998mb}
\begin{equation}\label{ini-sto}
Q_i=\frac{3\sigma_T}{4\pi\sigma_B}\sqrt{\frac{2\pi}{15}} \textrm{Re}\, a_{22},\qquad U_i=-\frac{3\sigma_T}{4\pi\sigma_B}\sqrt{\frac{2\pi}{15}} \textrm{Im}\, a_{22},
\end{equation}
where $\sigma_T$ is the Thomson scattering cross section, $\sigma_B$ is the cross sectional area of the scattered light and $a_{22}$ is the second multipole coefficient used in expanding the incident photon intensity in spherical harmonics $Y_{lm}$. We have intentionally labeled with $i$ the values of $Q$ and $U$ at the decoupling time and use them as the initial conditions in \eqref{sol-Q}-\eqref{sol-V}. However, as it has been well studied in the literature, Thomson scattering does not generate circular polarization and in the case of CMB this fact is confirmed since equation which governs evolution of $V$ parameter due to Thomson scattering has no source term \cite{Kosowsky:1994cy}. Consequently in this section we assume that at decoupling $V_i=0$.

In the second place we may note from \eqref{sol-Q}-\eqref{sol-U} that due to magneto-optic effects, light during its propagation contemporary has its polarization plane rotated and there is generation of phase shift between photon states $A_+$ and $A_\times$. This behavior is well known in optics and the medium which induces such effects is usually referred as rotated retarder, namely it rotates the polarization plane and generates ellipticity at the same time. We may note that for $V_i=0$ the expressions for $Q$ and $U$ are the same as in the case of solely Faraday effect taken into account. Moreover, in case when light is linearly polarized, $\Delta M(T)=0$ the Faraday effect alone does not generate circular polarization. Also we may note the contribution of the Hubble friction term to the Stokes parameters, namely $(T/T_i)^3$ that for convenience reasons we putted it on the r. h. s. of  \eqref{sol-I}-\eqref{sol-V}. The \emph{effective} scaling of the Stokes parameters due to universe expansion is not $(T/T_i)^3$ but $(T/T_i)^2$ since the Stokes parameters contain an intrinsic scaling of $(T/T_i)$. This is due to the fact that in the WKB approximation the fields have an intrinsic normalization of $1/\sqrt{\omega(t)}$ with $\omega(t)$ being photon/pseudoscalar energy. We did not show explicitly this factor for simplicity which gives a multiplicative factor to fields proportional to $T^{1/2}$ and to the Stokes parameters proportional to $T$. Since the scaling factor due to universe expansion is common to all Stokes parameters and because we are mostly interested in their ratio or expressions that contain their ratio, such as polarization degrees $P_{L, C}$, this term eventually cancels out. For example, the degree of linear polarization of the CMB remains constant during universe expansion to first order in $\epsilon$
\begin{equation}\nonumber
P_L(T)=\frac{\sqrt{Q^2(T)+U^2(T)}}{I(T)}=\sqrt{Q_i^2+U_i^2}=P_L(T_i),
\end{equation}
where we took for simplicity $I_i=1$ and $V_i=0$ in \eqref{sol-Q}-\eqref{sol-U}. The total rotation angle of linear polarization of the CMB due to the Faraday effect is given by $ \psi_F(T)=F(T)/2$. The contribution of CM and vacuum polarization effects to linear polarization does not appear to the first order in $\epsilon$. Their contribution appears only to second order in $\epsilon$ but for our purposes only expansion to first order is important in this section.

Apart the fact that the polarization plane of the CMB is rotated due to the Faraday effect, another interesting effect is the generation of circular polarization with non zero Stokes parameter $V(T)$. Even in the case when there is not circular polarization at the decoupling time, it is generated afterwards due to vacuum polarization and CM effects. In the case of vanishing $V_i$ we have
\begin{equation}
(T_i/T)^3V(T) =-Q_i\int_{T}^{T_i} \epsilon\,\tilde G(T^\prime)\sin[F(T^\prime)]dT^\prime-U_i\int_{T}^{T_i} \epsilon\,\tilde G(T^\prime)\cos[F(T^\prime)]dT^\prime\label{sol-V1}.
\end{equation}
Based on \eqref{sol-V1}, in this section we concentrate mostly in calculation of degree of circular polarization of the CMB in cases of vacuum polarization and CM effects. Their contribution is included in the term $\Delta M(T)$ where $\Delta M(T)=\Delta M_\textrm{CM}(T)+\Delta M_\textrm{QED}(T)$.

In both terms on the r. h. s. of \eqref{sol-V1} enters the function $F(T)$ which represents the effect of the Faraday effect. To have an analytic expression for $F(T)$ we need first the expression for $M_F(T)$ which is given by one of the off-diagonal terms of $\Pi^{ij}$. The Faraday effect is induced by the longitudinal component of the magnetic field with respect to $\bs k$, namely by $\bs B_L=\bs B_e \cos(\Phi)$. Consequently, linearly polarized electromagnetic wave propagating along the direction of the external magnetic field, has its polarization plane rotated with an angle proportional to $B_L$.  This occurs because the right and left handed indexes of refraction $n_R$ and $n_L$ are different from each other, which make possible mixing between linearly polarized states $A_+$ and $A_\times$. The expression for the Faraday term is given by
\begin{equation}\nonumber
M_F=|\Pi^{12}|/2\omega=\frac{\omega_\textrm{pl}^2\omega_c\cos(\Phi)}{2(\omega^2-\omega_c^2)},
\end{equation}
where $\omega_\textrm{pl}^2=4\pi\alpha n_e/m_e$ is the plasma frequency, $n_e$ is the free electron number density, $m_e$ is the electron mass and we used $k\simeq\omega$ for photons. Here $\omega_c=e B_e/m_e$ is the cyclotron frequency with $e$ being the electron charge. During propagation of the electromagnetic wave in a magnetized medium, the wave polarization remains unchanged for initial linearly polarized wave, but the linear polarized states $A_+$ and $A_\times$ propagate with a new index of refraction $\Delta n_F$ in the medium which is given by $\Delta n_F=n_R-n_L$.

The last thing that remains to calculate is the expression for the Hubble parameter which enters $F(T)$ in \eqref{fa-F} and in $\epsilon\tilde G(T)$. In general, its expression in the case of zero spatial curvature ($\kappa=0$) is given by
\begin{equation}\nonumber
H(T)=H_0\left(\Omega_\Lambda+\Omega_M(T/T_0)^3+\Omega_R(T/T_0)^4\right)^{1/2},
\end{equation}
where $H_0$ is the Hubble parameter at the present epoch, $H_0=H(T_0)$, $\Omega_\Lambda$ is the density parameter of the vacuum energy, $\Omega_M$ is the matter density parameter and $\Omega_R$ is the density parameter of relativistic particles. According to the Planck collaboration \cite{Ade:2013zuv}, values of density parameters of nonrelativistic matter and vacuum energy are respectively $h_0^2\Omega_M=0.12$ and $\Omega_\Lambda=0.68$ with $h_0=0.67$. The density parameter of relativistic particles it is straightforward to calculate, $\Omega_R=4.15\times 10^{-5} h_0^{-2}$ which includes the contribution of photons and three neutrino species assumed to be nearly massless. The contribution of the external magnetic field to the energy density budget of relativistic fields can be safely neglected since its energy density is $\rho_B(T_0)\simeq 10^{-7}(B_0/\textrm{nG})^2\rho_\gamma(T_0)$.

\subsection{Vacuum polarization in external magnetic field}
\label{sec:4.1}

Having the expressions for the Faraday term $M_F(T)$ and $H(T)$, we have almost all necessary ingredients to calculate the degree of circular polarization\footnote{The degree of circular polarization as discussed in Appendix \ref{app:1} is defined as $P_C=|V|/I$. Since we choose $I_i=1$ and because the scaling term due to universe expansion cancels out, $P_C=|V|$. Consequently in this section we calculate only $V(T)$ and take its absolute value if it is a negative quantity.} at present time, $V(T_0)$. Vacuum polarization and CM effects are responsible for generation of circular polarization. 
They are induced by the transverse component of the external magnetic field, $B_T$. In both effects the linear polarization indexes of refraction, $n_+$ and $n_\times$ are different from each other. Contrary to the Faraday effect which has its index of refraction proportional to $B_L$, vacuum polarization and CM effects have their indexes of refraction proportional to $B_T^2$.

\begin{figure*}[htbp!]
\begin{center}
\includegraphics[scale=0.8]{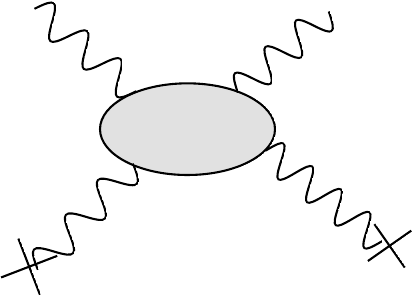}
\caption{Vacuum polarization in an external magnetic field. The cross vertexes denote external magnetic field and the wavy lines denote photons. In calculating the polarization tensor for the vacuum polarization, only the contribution of electron/positron loop is included.}
\label{fig:Fig2}
\end{center}
\end{figure*}

In this section we consider the contribution of vacuum polarization\footnote{Vacuum polarization in external magnetic field is a non linear QED effect which Lagrangian density is given by the Euler-Heisenberg term $\mathcal L_{EH}=\frac{\alpha^4}{90m_e^4}\left[(F_{\mu\nu}F^{\mu\nu})^2+\frac{7}{4}(F_{\mu\nu}\tilde F^{\mu\nu})^2\right]$.} to $V(T)$ separately from the CM effect which will be considered in the next section. Vacuum polarization\footnote{Vacuum polarization considered in this paper is due to interaction of CMB photons with an external magnetic field which is different from free CMB photon-photon scattering studied in Ref. \cite{Sawyer:2012gn}. } occurs (not only) in the presence of an external magnetic field due to creation of electron/positron pair from the vacuum, see Fig. \ref{fig:Fig2}. The expressions of elements of photon polarization tensor\footnote{The diagonal terms of the polarization tensor include the contribution of plasma effects, vacuum polarization and CM effect. Since plasma effects are the same for $A_+$ and $A_\times$ and because in this section and in the next we calculate, $\Delta M=M_+-M_\times$, the plasma term cancels out.} corresponding to the states $A_+$ and $A_\times$ in case of vacuum polarization, for slowly varying external magnetic field in space and time over the Compton wavelength, are respectively given by \cite{Brezin:1971nd}
\begin{equation}\label{qed-pol}
\Pi_\textrm{QED}^{22}=-4\kappa\,\omega^2\sin^2(\Phi), \quad \Pi_\textrm{QED}^{11}=-7\kappa\,\omega^2\sin^2(\Phi),
\end{equation}
where $\kappa=(\alpha/45\pi)(B_e/B_c)^2$ and $B_c=m_e^2/e$ is the critical magnetic field. Using \eqref{qed-pol} and definitions of $M_+$ and $M_\times$ we get
\begin{equation}\label{del-qed}
\Delta M_\textrm{QED}=-\frac{3}{2}\kappa\,\omega\sin^2(\Phi).
\end{equation}

Now using the definition of $\kappa$ and taking into account that the photon energy scales with the temperature as $\omega=\omega_0(T/T_0)$ and assuming magnetic field flux conservation in the cosmological plasma with $B_e(T)=B_e(T_0)(T/T_0)^2$ we get
\begin{eqnarray}
G(T) &= & \epsilon_\textrm{QED} \tilde G(T)= -8.12\times 10^{-14}\left(\frac{\nu_0}{\textrm{Hz}}\right)\left(\frac{B_{e 0}}{\textrm{G}}\right)^2 \left(\frac{T}{T_0}\right)^{5/2}\sin^2(\Phi) \qquad (\textrm{K}^{-1})\nonumber,\\
F(T) &=& 8.71\times 10^{25}\cos(\Phi)\left(\frac{\textrm{Hz}}{\nu_0}\right)^2\left(\frac{B_{e 0}}{\textrm{G}}\right)\int_{T}^{T_i} X_e(T^\prime)\left(\frac{T^\prime}{T_0}\right)^{1/2}d T^\prime\quad (\text{K}^{-1})\label{F-1},
\end{eqnarray}
where $T_0$ is the CMB temperature today and $\omega_0=2\pi\nu_0$ with $\nu_0$ being the CMB frequency at present. Deriving \eqref{F-1} we used the fact that $\omega_c\ll \omega$ in the Faraday term, expressed the free electron number density $n_e(T)$ which enters the plasma frequency as $n_e(T)=0.76\, n_B(T_0) X_e(T)(T/T_0)^3$ with $X_e(T)$ being the ionization fraction of free electrons and $n_B(T_0)$ is the baryon number density at present. Moreover, we assumed for simplicity that only non relativistic matter contributes to the Hubble parameter, namely $H(T)\simeq H_0 \sqrt{\Omega_M}(T/T_0)^{3/2}$. Until now we did not give any explicit expression for the parameter $\epsilon_\textrm{QED}$. This parameter can be extracted immediately from $G(T)$ in \eqref{F-1} and it is given by
\begin{equation}\nonumber
\epsilon_\textrm{QED} \equiv -8.12\times 10^{-14}\left(\frac{\nu_0}{\textrm{Hz}}\right)\left(\frac{B_{e 0}}{\textrm{G}}\right)^2T_0^{-5/2} \sin^2(\Phi)
\qquad (\textrm{K}^{-1}).
\end{equation}

With expressions of $F(T), G(T)$ and $\epsilon_\textrm{QED}$ we have all necessary quantities to calculate $V(T)$ in \eqref{sol-V1}. Let us start with the first term on the r. h. s. of  \eqref{sol-V1} which has the following dependence on the temperature
\begin{equation}\label{int-eq}
\int_{T}^{T_i} \epsilon_\textrm{QED}\tilde G(T^\prime)\sin[F(T^\prime)]\,dT^\prime=\epsilon_\textrm{QED}\, \int_{T}^{T_i} T^{\prime 5/2}\sin\left[\rho\,\int_{T^\prime}^{T_i} X_e(T^{\prime\prime})\,T^{\prime\prime 1/2} d T^{\prime\prime}\right] d T^\prime,
\end{equation}
where we have defined 
\begin{equation}\nonumber
\rho \equiv 8.71\times 10^{25}\cos(\Phi)\left(\frac{\textrm{Hz}}{\nu_0}\right)^2\left(\frac{B_{e 0}}{\textrm{G}}\right)T_0^{-1/2}\qquad \textrm{K}^{-1}.
\end{equation}
As one would expect there is an integration in $X_e(T)$ on the r. h. s. of expression \eqref{int-eq} that complicates the situation quite a lot. Indeed, there is no known analytic expression for $X_e(T)$ which in general satisfies a complicated differential equation, see Ref. \cite{Weinberg:2008zzc} for details. At the temperature $T\simeq 3000$ K numerical solution of the equation satisfied by $X_e(T)$, shows that ionization fraction is $X_e\simeq 0.13$ and drops down to $X_e\simeq 2\times 10^{-2}$ at the temperature $T=2000$ K. When the temperature is about 200 K it drops down to $X_e\simeq 2.7\times 10^{-4}$ and remains almost constant afterwards if no reionization epoch is assumed. In this work we use the solution for $X_e(T)$ given in Ref. \cite{Weinberg:2008zzc} and interpolate it with $X_e\simeq 1$ for $T\lesssim 21.8$ K which corresponds to the period of end of the reionization epoch.

The integral in \eqref{int-eq} has analytic solution in terms of the incomplete Euler gamma functions, if the expression for $X_e(T)$ is constant\footnote{In principle one can obtain analytic solution for the integral \eqref{int-eq} by considering the average value of $X_e(T)$ at the post decoupling epoch as we have shown for some cases in Sec. \ref{sec:6}. However, we don't need to do it here because the vacuum polarization dominates in general the CM effect for $\nu_0> 10^{10}$ Hz. Consequently, we can get more accurate result by considering the numerical solution for $X_e$, expand the argument of sine function and integrate it numerically. }. However, one may observe that for $T\leq 2970$ K, for magnetic field strength $B_{e0}\leq 1$ nG and frequencies $\nu_0\geq 10^{10}$ Hz, the expression inside sine function is less than unity, namely $F(T)< 1$. In this case one can use series expansion and consider only the first term. Consequently we obtain
\begin{equation}\label{int-eq-2}
\int_{T}^{T_i} \epsilon_\textrm{QED}\tilde G(T^\prime)\sin[F(T^\prime)]\,dT^\prime\simeq \epsilon_\textrm{QED}\,\rho \, \int_{T}^{T_i} T^{\prime 5/2}\int_{T^\prime}^{T_i} X_e(T^{\prime\prime})\,T^{\prime\prime 1/2} d T^{\prime\prime} d T^\prime.
\end{equation}
We are interested in calculating the integral in \eqref{int-eq-2} at $T=T_0$ and numerical calculation gives
\begin{equation}\nonumber
\int_{T_0}^{T_i} T^{\prime 5/2}\int_{T^\prime}^{T_i} X_e(T^{\prime\prime})\,T^{\prime\prime 1/2} d T^{\prime\prime} d T^\prime\simeq 5.17\times 10^{14}\,\qquad (\textrm{K}^5),
\end{equation}
and expression \eqref{int-eq-2} becomes 
\begin{equation}\label{int-eq-3}
\int_{T_0}^{T_i} \epsilon_\textrm{QED}\tilde G(T^\prime)\sin[F(T^\prime)]dT^\prime= 5.17\times 10^{14} \epsilon_\textrm{QED}\,\rho \,\qquad (\textrm{K}^5).
\end{equation}

Now it remains to calculate the second term on the r. h. s. of \eqref{sol-V1}. Based on the same arguments as we did above for the first term, the argument of cosine function is smaller than unity and we can write
\begin{equation}\nonumber
\int_{T_0}^{T_i} \epsilon_\textrm{QED}\,\tilde G(T^\prime)\cos[F(T^\prime)]dT^\prime\simeq \frac{2}{7}\,\epsilon_\textrm{QED}\,\left(T_i^{7/2}-T_0^{7/2}\right).
\end{equation}
The value of $V$ at present time would be for $T_i=2970$ K (the CMB temperature at the redshift $1+z=1090$ corresponding to the decoupling time) and $T_0=2.725$ K
\begin{equation}\label{v-today}
V_0(\nu_0, B_0, \Phi)\simeq 1.8\times 10^{26}\,\sin^2(\Phi)\cos(\Phi)\left(\frac{\textrm{Hz}}{\nu_0}\right)\left(\frac{B_{e 0}}{\textrm{G}}\right)^3\,Q_i+2.7\times 10^{-3}\,\sin^2(\Phi)\left(\frac{\nu_0}{\textrm{Hz}}\right)\left(\frac{B_{e 0}}{\textrm{G}}\right)^2\,U_i.
\end{equation}
If for example we take $\nu_0=30$ GHz and $B_{e0}=1$ nG, we get
\begin{equation}\label{tot-v}
V_0(\Phi)\simeq 6\times 10^{-12}\,\sin^2(\Phi)\cos(\Phi)\,Q_i+8.1\times 10^{-11}\,\sin^2(\Phi)\,U_i.\end{equation}
We have checked that numerical values derived from expression \eqref{v-today}, perfectly agree with numerical solutions in the case when one assumes $H\simeq H_0\sqrt{\Omega_M}(T/T_0)^{3/2}$. The difference between numerical solutions with total $H$ and semi-analytic solutions with only matter contribution to $H$, is that in the former case numerical solutions are in general smaller by a factor less than $\sqrt{2}$ with respect to the latter case. We may note that the second term in \eqref{tot-v} is proportional to the frequency and for higher values of $\nu_0$, $V_0$ increases linearly with $\nu_0$. 

So far, we have derived our results in the case when $\Phi\neq \pi/2$, which allowed us to find perturbative solution for the Stokes vector $S(T)$. In the case when the magnetic field is transverse, this approximation is not valid anymore since $M_F(T)\rightarrow 0$ for $\Phi\rightarrow \pi/2$. However, if $\Phi=\pi/2$ it is not necessary to work with the perturbative approach since the equation for the Stokes vector simplifies significantly. Indeed, for $\Phi= \pi/2$ there is only mixing between the Stokes parameters $U$ and $V$. The solution of equations of motion for the Stokes parameters $U$ and $V$, in transverse magnetic field are immediate and read
\begin{equation}\label{trans-par}
Q(T) = Q_i,\quad U(T) =  \cos\left[\mathcal G(T)\right]U_i+\sin\left[\mathcal G(T)\right]V_i,\quad
V(T) = -\sin\left[\mathcal G(T)\right] U_i+\cos\left[\mathcal G(T)\right]V_i,
\end{equation}
where we have defined $\mathcal G(T) \equiv \int_T^{T_i} G(T^\prime) dT^\prime$.

In case of $V_i=0$, one would get for $V$
\begin{equation}\nonumber
V(T) =-\sin\left[\mathcal G(T)\right] U_i.
\end{equation}
In order to estimate $V(T)$ at present time for $\Phi=\pi/2$, we first must calculate $\mathcal G(T)$ in the argument of sine function. Consequently, we get
\begin{equation}\nonumber
\mathcal G(T)=-2.32\times 10^{-14}T_0^{-5/2}\left(\frac{\nu_0}{\textrm{Hz}}\right)\left(\frac{B_{e 0}}{\textrm{G}}\right)^2 \left(T_i^{7/2}-T^{7/2}\right)\quad (\text{K}^{-1}),
\end{equation}
 and the value of $V$ at $T=T_0$ is
 \begin{equation}\label{transverse-b}
V_0(\nu_0, B_{e0})\simeq \sin\left[2.7\times 10^{-3}\,\left(\frac{\nu_0}{\textrm{Hz}}\right)\left(\frac{B_{e 0}}{\textrm{G}}\right)^2\right]\,U_i.
\end{equation}
In general for a wide range of the parameters $\nu_0$ and $B_{e0}$ in \eqref{transverse-b}, the argument of sine function is much less than unity and one can replace to first order the sine with its argument. If we take for example the values $\nu_0=100$ GHz and $B_{e0}=1$ nG we get
\begin{equation}\nonumber
V_0\simeq 2.7\times 10^{-10}\,U_i.
\end{equation}
If the magnetic field is 10 nG we would get $V_0\simeq 2.7\times 10^{-8}\,U_i$. It is worth to note that in case of transverse magnetic field, the vacuum polarization induces also a rotation of the polarization plane. Indeed, as can be inferred from \eqref{trans-par}, the rotation angle of the polarization plane is given by
\begin{equation}\label{qed-lin-tans}
\tan[2\psi(T)]=\tan[2\psi(T_i)] \cos\left[\mathcal G(T)\right],\end{equation}
which in general is a very small quantity for vacuum polarization.

Until now we kept the dependence on $\Phi$ explicit in \eqref{v-today} but it is more convenient to average over all possible orientations of $\bs B_e$ relative to $\bs k$. We must note that \eqref{v-today} has been derived by assuming $\Phi\neq \pi/2$ which allowed us to find perturbative solution for the Stokes vector $S(T)$. However, we may note that in the limit $\Phi\rightarrow \pi/2$, the first term in \eqref{v-today} goes to zero while the second term coincides with the argument of sine function in \eqref{transverse-b} which has been found exactly. This fact allows us, to find the following expression for the rms of $V_0$ in \eqref{v-today}, which for $F(T)< 1$ is given by
\begin{equation}\label{rms-V0}
\langle V_0^2(\nu_0, B_{e0})\rangle^{1/2}\simeq \left[2\times 10^{51}\left(\frac{\textrm{Hz}}{\nu_0}\right)^2\left(\frac{B_{e 0}}{\textrm{G}}\right)^6\,Q_i^2+2.73\times 10^{-6}\,\left(\frac{\nu_0}{\textrm{Hz}}\right)^2\left(\frac{B_{e 0}}{\textrm{G}}\right)^4\,U_i^2\right]^{1/2}.
\end{equation}
Assuming for example, $B_{e0}=1$ nG, $Q_i\simeq U_i$ we get for $\nu_0=30$ GHz
and $\nu_0=700$ GHz respectively $\langle V_0^2\rangle^{1/2}\simeq 4.96\times 10^{-11}\,Q_i$ and $\langle V_0^2\rangle^{1/2}\simeq 1.15\times 10^{-9}\,Q_i$. We may note from \eqref{rms-V0}, that biggest contribution comes for values of $\Phi\rightarrow\pi/2$ or transverse fields and for higher values of $B_{e0}$, the rms of $V_0$ is bigger.

Another interesting case is when the arguments of sine and cosine functions in \eqref{trans-par} are equal to $\pi/2$, namely $\mathcal G(T)=\pi/2$. This condition is fulfilled when
\begin{equation}\label{relation-om-b}
\left(\frac{\nu_0}{\textrm{Hz}}\right)=581.3\,\left(\frac{\textrm{G}}{B_{e0}}\right)^2.
\end{equation}
When this condition is met, we would have $V(T)=U_i$. However, in order for condition \eqref{relation-om-b} to be fulfilled, the value of $\nu_0$ must be much far beyond the present observed CMB spectrum for reasonable values of $B_{e0}$.

\subsection{The Cotton-Mouton effect}
\label{sec:4.3}

As briefly mentioned in the previous sections, the CM effect is a birefringence effect which is induced in a medium in presence of transverse external magnetic field and which generates elliptic polarization. This effect has been studied and experimented in gases and liquids where limits on the CM constant $C_\textrm{CM}$ are set or its value is established. However, the CM effect does not only manifest in gases and liquids but also in plasma\footnote{I learned only recently about the CM effect in plasma.}. The theory of this phenomena is studied to some extend classically and also quantum mechanically and for a discussion on this mechanism see Ref. \cite{Born65}.

After the decoupling epoch, the ionization fraction of free electrons rapidly dropped down to an almost constant value of $X_e\simeq 2.7\times 10^{-4}$ and later again it increased to $X_e\simeq 1$ at the reionization epoch. On the other hand almost all baryons would bind together to form the light elements such as atomic hydrogen and helium etc. This state of mixed hydrogen and helium gas (plus a small fraction of other light elements) with the electron plasma continued coexisting till the start of the reionization epoch. In order to study the impact of CM effect in the generation of CMB polarization, we need the value of $\Delta M_\textrm{CM}$ for hydrogen and helium gas and for the electron plasma.

Theory of CM effect in gases has been extensively studied in the literature and for a review on the subject see Ref. \cite{Rizzo-97} and references there. In case of gases, theoretical calculations give the following expression for the difference in index of refraction $\Delta n_\textrm{CM}^\textrm{gas}$ \cite{Rizzo-97}
\begin{equation}\label{CM-index}
\Delta n_\textrm{CM}^\textrm{gas}=\pi B_e^2\sin^2(\Phi)\,n^\textrm{gas}\,\Delta\eta/(4\pi\varepsilon_0),
\end{equation}
where $n^\textrm{gas}$ is the gas number density and $\Delta\eta$ is called the hypermagnetizability anisotropy. Here it is assumed that $n^\textrm{gas}$ obeys the perfect gas law or its closely related ideal gas law. In general $\Delta\eta$ will depend on the type of gas and on the incident energy of the electromagnetic radiation. Quite often the CM constant is also defined through the relation
\begin{equation}\label{del-cm}
\Delta n_\textrm{CM}^\textrm{gas} =C_\textrm{CM}\, \lambda\, B_e^2\,\sin^2(\Phi),
\end{equation}
where $\lambda$ is the wave length of the incident electromagnetic wave (not to be confused with photon helicity state). By comparing \eqref{del-cm} with \eqref{CM-index} we get the following expression for 
$C_\textrm{CM}=\pi\,\Delta\eta\,n^\textrm{gas}/\lambda.$

In our case, we are interested in only the magnetic hypermagnetizability of hydrogen and helium gases since these elements are the most abundant ones and neglect the contribution of the other light elements. Following Ref. \cite{Rizzo-97}, theoretical values of $\Delta\eta$ in the limit of zero incident photon momentum, gas temperature $T_\textrm{gas}=273.15$ K and gas pressure $P_\textrm{gas}=$ 1 atm are respectively given by $\Delta\eta_\textrm{H}=13.33$\, au and $\Delta\eta_\textrm{He}=1.06\,\text{au}$ where $1$ au of $\eta$ is $\simeq 2.682\times 10^{-44} (4\pi\varepsilon_0)$ G$^{-2}$ cm$^{-3}$ \cite{Rizzo-97} with $\varepsilon_0=1$ in the rationalized Lorentz-Heaviside system. Consequently we get
\begin{equation}\nonumber
\Delta M_\textrm{CM}^\textrm{gas}\simeq\pi\omega B_e^2\sin^2(\Phi)\,(Y_\textrm{H}\Delta\eta_\textrm{H}+Y_\textrm{He}\Delta\eta_\textrm{He})\,n_B/(4\pi\varepsilon_0),
\end{equation}
where $Y_\textrm{H}$, $Y_\textrm{He}$ are respectively the primordial abundances of atomic hydrogen and helium. Assuming that $\Delta\eta$ does not change significantly in the frequency range corresponding to the CMB after the decoupling epoch, we get
\begin{equation}\label{CM-gas}
\Delta M_\textrm{CM}^\textrm{gas}(T)= 1.2\times 10^{-60}\,\ \left(\frac{\nu_0}{\textrm{Hz}}\right)\left(\frac{B_{e 0}}{\textrm{G}}\right)^2\,\sin^2(\Phi)\left(\frac{T}{T_0}\right)^8 \qquad (\textrm{K}).
\end{equation}

The contribution of the electron plasma to the CM effect enters the diagonal elements of the polarization tensor in magnetized plasma, $\Pi^{11}$ and $\Pi^{22}$. The difference with respect to the Faraday effect is that CM effect is quadratic in the amplitude of transverse magnetic field and one would expect that for typical values of the cosmic magnetic field, its magnitude would be much weaker than the Faraday effect. In general, for a magnetized plasma the contribution of the CM effect to the photon polarization tensor is given by \cite{D'Olivo:2002sp}
\begin{equation}\label{cm-pol-el}
\Pi_\textrm{CM}^{11}=\frac{\omega^2\,\omega_\textrm{pl}^2}{\omega^2-\omega_c^2}-\frac{\omega_\textrm{pl}^2\,\omega_c^2}{\omega^2-\omega_c^2}\sin^2(\Phi), \quad \Pi_\textrm{CM}^{22}=\frac{\omega^2\,\omega_\textrm{pl}^2}{\omega^2-\omega_c^2}.
\end{equation}
We may note that in \eqref{cm-pol-el}, the CM term appears only in $\Pi^{11}$ (the second term) while it does not appear in $\Pi^{22}$. This is due to the fact that we have chosen since the beginning the transverse part of $B_e$ along the $x$ axis with no $y$ component. Using the definition of $\Delta M$, for the CM effect in magnetized plasma, we get
\begin{equation}\nonumber
\Delta M_\textrm{CM}^\textrm{pl}=-\frac{\omega_\textrm{pl}^2\omega_c^2}{2\omega(\omega^2-\omega_c^2)}\,\sin^2{\Phi}.
\end{equation}
In case of CMB, we have that photon frequency is much bigger than cyclotron frequency and we can approximate $\omega^2-\omega_c^2\simeq \omega^2$. Consequently, we get
\begin{equation}\label{CM-pl}
\Delta M_\textrm{CM}^\textrm{pl}=-2.82\times 10^3\,\ \left(\frac{\textrm{Hz}}{\nu_0}\right)^3\left(\frac{B_{e 0}}{\textrm{G}}\right)^2\,X_e(T)\left(\frac{T}{T_0}\right)^4\sin^2(\Phi) \qquad (\textrm{K}).
\end{equation}
If we compare \eqref{CM-pl} with \eqref{CM-gas}, we may observe that for the parameter space of magnetic field amplitude $B_{e0}$ and photon frequency $\nu_0$ of interest, the contribution of hydrogen and helium gases to the CM effect is much smaller than the contribution of electron plasma. Therefore from now we will neglect the gas contribution to the CM effect.

Now we can calculate the contribution of CM effect to $V(T)$ in the same way as we did for the vacuum polarization. In case when $\Phi\neq \pi/2$ and $F(T)< 1$ at post decoupling epoch, we have
 \begin{eqnarray}\label{cm-num-1}
\int_{T_0}^{T_i} G(T^\prime)\sin[F(T^\prime)]dT^\prime &\simeq &\epsilon_\textrm{CM}\,\rho\,\int_{T_0}^{T_i} T^{\prime 3/2}\,X_e(T^\prime)\int_{T^\prime}^{T_i} X_e(T^{\prime\prime})\,T^{\prime\prime 1/2} d T^{\prime\prime} d T^\prime,\nonumber\\
&=& 3.46\times 10^9\,\epsilon_\textrm{CM}\,\rho\,\qquad (\textrm{K}^{4}),
\end{eqnarray}
where in the second term in \eqref{cm-num-1} numerical integration has been used and defined $G(T)=\epsilon_\textrm{CM}\tilde G(T)$ with $\epsilon_\textrm{CM}$
\begin{equation}\nonumber
\epsilon_\textrm{CM}=-1.21\times 10^{32}\,\ \left(\frac{\textrm{Hz}}{\nu_0}\right)^3\left(\frac{B_{e 0}}{\textrm{G}}\right)^2T_0^{-3/2}\,\sin^2(\Phi) \qquad (\textrm{K}^{-1}).
\end{equation}
The second term that enters the r. h. s. of \eqref{sol-V1} is given by
\begin{equation}\nonumber
\int_{T_0}^{T_i} G(T^\prime)\cos[F(T^\prime)]dT^\prime\simeq \epsilon_\textrm{CM}\,\int_{T_0}^{T_i} T^{\prime 3/2}\,X_e(T^\prime)\,dT^\prime=4.45\times 10^6\,\epsilon_\textrm{CM}\,\qquad (\textrm{K}^{5/2}).
\end{equation}
The total expression for the degree of circular polarization $V(T)$ at present time is given by
\begin{equation}\label{CM-V}
V_0(\omega_0, B_{e0}, \Phi)\simeq -3.46\times 10^9\,\epsilon_\textrm{CM}\,\rho\,Q_i-4.45\times 10^6\,\epsilon_\textrm{CM}\,U_i.
\end{equation}
Now we can put some numbers in \eqref{CM-V} in order to estimate $V_0$. For example in the case when $\nu_0=30$ GHz and $B_{e0}=1$ nG we get 
\begin{equation}
V_0=2\times 10^{-13}\sin^2(\Phi)\cos(\Phi)\,Q_i+4.43\times 10^{-12}\sin^2(\Phi)\,U_i.\end{equation}

The case when the magnetic field is completely transverse, $\Phi=\pi/2$ is treated in the same way as in the case of vacuum polarization. What we need is to calculate $\mathcal G(T)$ in the case of CM effect, which in most cases is $\ll 1$. We remind that expressions for the Stokes parameters in case of transverse magnetic field are found exactly without using perturbation theory and are given in \eqref{trans-par}. In case when $V_i=0$, from \eqref{trans-par} we get 
\begin{equation}
Q_0 = Q_i,\quad U_0 \simeq U_i,\quad V_0 \simeq 1.2\times 10^{38}\left(\frac{\textrm{Hz}}{\nu_0}\right)^3\left(\frac{B_{e0}}{\textrm{G}}\right)^2\,U_i,\label{cm-in-cube}
\end{equation}
where we kept only the first order term in $\mathcal G(T)\ll1$. The most interesting fact, is the relation $V_0\propto \nu_0^{-3}$ in \eqref{cm-in-cube}. If we consider for example $B_{e0}=1$ nG and $\nu_0=10^8$ Hz we get $V_0\simeq 1.2\times 10^{-4}U_i$ while for $\nu_0=10^9$ Hz we get $V_0\simeq 1.2\times 10^{-7}U_i$. In principle one can also calculate the rms of $V_0$ for the CM effect, as we did in case of vacuum polarization, but it is not that easy. Indeed, expression \eqref{CM-V} has been derived in the approximation when $F(T)< 1$, which assuming that $B_{e0}\leq 1$ nG, it is satisfied for $\nu_0>10^{10}$ Hz. However, the biggest contribution in $V_0$ comes from the low frequency part as we saw for the case when $\Phi=\pi/2$. Instead of looking for analytic solution even when $F(T)< 1$ is not satisfied and after estimate the rms of $V_0$, one possible way to circumvent this situation, is to note that rms of $V_0$ is bigger for values of $\Phi\rightarrow \pi/2$. Indeed, we have checked the numerical solution and found that value of $V_0$ for $\Phi\neq \pi/2$, fixed $B_{e0}$ and $\nu_0$ is much smaller than that of $\Phi=\pi/2$. Consequently, one can approximate to very good accuracy the rms of $V_0$ with its value at $\Phi=\pi/2$.  

The low frequency part of the CMB, $\nu_0\sim 10^8$ Hz for $\Phi=\pi/2$ is very interesting since significant rotation of the polarization plane occurs. Indeed, if one keeps $\mathcal G(T)\ll 1$ up to second order in $U(T)$ in expression \eqref{trans-par} and assuming that at decoupling time $Q_i\simeq -U_i$, one gets for the rotation angle the following expression
\begin{equation}
\delta\psi(T_0)\simeq 1.8\times 10^{75}\left(\frac{\textrm{Hz}}{\nu_0}\right)^6\left(\frac{B_{e0}}{\textrm{G}}\right)^4,
\end{equation}
where we wrote $\psi(T)\simeq \psi(T_i)+\delta\psi(T)$ with $|\delta\psi|\ll 1$ and used \eqref{qed-lin-tans}. If we consider $B_{e0}=1$ nG, $\nu_0\simeq 10^8$ Hz we get $\delta\psi(T_0)\simeq 1.8\times 10^{-9}$ rad, for $B\simeq 10^{-8}$ G and same frequency we would get $\delta\psi(T_0)\simeq 1.8\times 10^{-5}$ rad and for $B=10^{-7}$ G we would get $\delta\psi(T_0)=0.18$ rad.

\section{Three state density matrix and generalized Stokes parameters}
\label{sec:5}

In Sec. \ref{sec:4} we derived temperature (or time) evolution of the Stokes parameters in case when only Faraday effect, vacuum polarization and CM effects were included in the equations of motion of the density matrix. However, there still remain one effect left which is the photon-pseudosocalar mixing in magnetic field. Including this new effect, the equations of motion of the density matrix become more involved and instead of four equations which had in Sec. \ref{sec:4}, now we have nine of them. This can be verified by inserting the mixing matrix $M$ and the damping matrix $D$ in Eq. \eqref{dens-eq-1}. After we get the following system of differential equations
\begin{eqnarray}\label{ax-system}
\dot \rho_{11} &=& -M_{F}(\rho_{12}+\rho_{21})-3H\rho_{11},\quad
\dot\rho_{12} = -M_{F}(\rho_{22}-\rho_{11})+i(M_+-M_\times)\rho_{12}- M_{\phi\gamma}\rho_{13}-3H\rho_{12},\nonumber\\
\dot\rho_{13} &=& -M_{F}\rho_{23}-i(M_\phi-M_+)\rho_{13}+M_{\phi\gamma}\rho_{12}-3H\rho_{13},\quad
\dot\rho_{21} = \dot\rho_{12}^{*},\nonumber\\
\dot\rho_{22} &=& M_{F}(\rho_{12}+\rho_{21})-M_{\phi\gamma}(\rho_{32}+\rho_{23})-3H\rho_{22},\quad 
\dot\rho_{23} = M_{F}\rho_{13}+i(M_\times-M_\phi)\rho_{23}+M_{\phi\gamma}(\rho_{22}-\rho_{33})-3H\rho_{23},\nonumber\\
\dot\rho_{31} &=& \dot \rho_{13}^{*},\quad
\dot\rho_{32} = \dot\rho_{23}^{*},\quad
\dot\rho_{33} = M_{\phi\gamma}(\rho_{23}+\rho_{32})-3H\rho_{33}\label{ax-system-1},
\end{eqnarray}
where the sign ($^*$) means complex conjugate of a C-number and each element of the mixing matrix $M$ depends on the cosmological time $t$, $B_e$ and $\nu$. We may observe that total intensity is diluted due to universe expansion only
\begin{equation}\nonumber
\dot \rho_{11}+\dot\rho_{22}+\dot\rho_{33}=-3H(\rho_{11}+\rho_{22}+\rho_{33}),
\end{equation}
which means that trace of the density matrix is not constant in time. 

The system of Eq. \eqref{ax-system} still is not in the desired form since the photon intensity, that in this section we denote with $I_\gamma$, is not a conserved quantity. As already discussed in Sec. \ref{sec:4}, this is due to the fact that we are dealing with an open system interacting with the background. Even in the case of other magneto-optic effects which we treated in Sec. \ref{sec:4} there is interaction with the background, but with the difference that these effects conserve the photon number with momentum $\bs k$. Since $I_\gamma$ is not conserved, it would be convenient to express the equations of elements of the density matrix in terms of generalized Stokes parameters\footnote{Here `generalized Stokes parameters' does not mean a generalization to the case of $n\in \bf{N}$ states but simply means going from the description of two state parameters to the three state parameters.}, that is extending the usual two state Stokes parameters to the case of three states.

The derivation of generalized Stokes parameters can be done in analogous way as one does with  usual Stokes parameters. As shown in Appendix \ref{app:1}, one can express the elements of the two dimensional density matrix in terms of the Stokes parameters and one can check from direct calculation that expression \eqref{op-density} can be written in terms of the Pauli matrices $\sigma_i$ as follows
\begin{equation}\nonumber
\rho=\frac{1}{2}\left(S_0\,\bs I_{2\times 2}+\sum_{i=1}^3S_i\sigma_i\right),
\end{equation}
where we recall that $\langle S_0\rangle=I_\gamma, \langle S_1\rangle=U, \langle S_2\rangle=V, \langle S_3\rangle=Q$ and $\bs I_{2\times 2}$ is the two dimensional identity matrix. The generalization of the usual two state Stokes parameters to the three state case can be done as follows
\begin{equation}\label{gen-stokes}
\langle \hat S_k\rangle:=\textrm{Tr}(\rho\lambda_k),
\end{equation}
where $\rho$ is the $3\times 3$ density matrix, $\hat S_{k}$ (for $k\geq 1$) are the generators of SU(3) group and $\lambda_{k}$ (for $k\geq 1$), $(k=0, 1,...8)$, are the so called Gell-Mann matrices 
\begin{eqnarray}\nonumber
\lambda_0 &=& \bs I_{3\times 3}=\left(
\begin{matrix}
  1 & 0 & 0\\
  0 & 1 & 0\\
  0 & 0 & 1\\
 \end{matrix}\right),\quad 
\lambda_1=\left(
\begin{matrix}
  0 & 1 & 0\\
  1 & 0 & 0\\
  0 & 0& 0\\
 \end{matrix}\right),\quad 
 \lambda_2=\left(
\begin{matrix}
  0 & -i & 0\\
  i & 0 & 0\\
  0 & 0& 0\\
 \end{matrix}\right),\quad
 \lambda_3=\left(
\begin{matrix}
  1 & 0 & 0\\
  0 & -1 & 0\\
  0 & 0 & 0\\
 \end{matrix}\right), \nonumber\\ 
 \lambda_4 &= &\left(
\begin{matrix}
  0 & 0 & 1\\
  0 & 0 & 0\\
  1 & 0 & 0\\
 \end{matrix}\right), \quad 
   \lambda_5=\left(
\begin{matrix}
  0 & 0 & -i\\
  0 & 0 & 0\\
  i & 0 & 0\\
 \end{matrix}\right),\quad
  \lambda_6=\left(
\begin{matrix}
  0 & 0 & 0\\
  0 & 0 & 1\\
  0 & 1 & 0\\
 \end{matrix}\right),\quad
  \lambda_7=\left(
\begin{matrix}
  0 & 0 & 0\\
  0 & 0 & -i\\
  0 & i & 0\\
 \end{matrix}\right),\quad
  \lambda_8 =  \frac{1}{\sqrt{3}}\left(
\begin{matrix}
  1 & 0 & 0\\
  0 & 1 & 0\\
  0 & 0 & -2\\
 \end{matrix}\right)\nonumber.
 \end{eqnarray}
Inserting $\lambda_k$ into expression \eqref{gen-stokes} we can get the explicit expressions for the generalized Stokes parameters. The first set of four Stokes parameters is given by \eqref{def-stokes} while the remaining set of parameters is given by
\begin{equation}\label{new-set}
S_4 = \rho_{13}+\rho_{31},\quad S_5=i(\rho_{13}-\rho_{31}),\quad S_6=\rho_{23}+\rho_{32},\quad S_7 =i(\rho_{23}-\rho_{32}),\quad S_8 = \frac{1}{\sqrt{3}}(\rho_{11}+\rho_{22}-2\rho_{33}).
\end{equation}
The corresponding Stokes operators to the set \eqref{new-set} in the basis $|A_+\rangle, |A_\times\rangle, |\phi\rangle$ are given by
\begin{align}
\hat S_4 &= |A_+\rangle\langle \phi| + |\phi\rangle\langle A_+|,\quad \hat S_5 = i(|A_+\rangle\langle \phi| - |\phi\rangle\langle A_+|),\quad \hat S_6 = |A_\times\rangle\langle \phi| + |A_\times\rangle\langle\phi|,\nonumber\\
\hat S_7 &= i\left(|A_\times\rangle\langle \phi| - |\phi\rangle\langle A_\times|\right),\quad \hat S_8 = \frac{1}{\sqrt{3}}(|A_+\rangle\langle A_+| + |A_\times\rangle\langle A_\times| -2|\phi\rangle\langle \phi|)\nonumber.
\end{align}

Having defined the generalized Stokes parameters, now we are at the position to parametrize three state density matrix in terms of them as follows
\begin{equation}\label{gen-rho}
 \rho = \frac{1}{2}\left(
\begin{matrix}
  \frac{2}{3} I+Q+\frac{1}{\sqrt{3}} S_8 & U-iV & S_4-iS_5\\
  U+iV & \frac{2}{3} I-Q+\frac{1}{\sqrt{3}} S_8 & S_6-iS_7\\
  S_4+iS_5 & S_6+iS_7 & \frac{2}{3} I-\frac{2}{\sqrt{3}} S_8\\
 \end{matrix}\right),
 \end{equation}
where $I$ is the total intensity which is given by $I=I_\gamma+I_\phi$. Using \eqref{gen-rho} we can write the system of Eqs. \eqref{ax-system} as follows
\begin{eqnarray}
\dot I_\gamma &=& -M_{\phi\gamma} S_6-3HI_\gamma, \quad \dot Q = -2M_F U+M_{\phi\gamma} S_6-3HQ,\quad 
\dot U = 2 M_F Q+(M_+-M_\times) V-M_{\phi\gamma} S_4-3HU,\nonumber\\
\dot V &=& -(M_+-M_\times)U-M_{\phi\gamma} S_5-3HV,\quad
\dot S_4 = -M_{F}S_6+(M_+-M_\phi)S_5+M_{\phi\gamma}U-3HS_4,\nonumber\\
\dot S_5 &=& -M_{F}S_7-(M_+-M_\phi)S_4+M_{\phi\gamma}V-3HS_5,\quad
\dot S_6 = M_{F}S_4+(M_\times-M_\phi)S_7+M_{\phi\gamma}(\sqrt{3}S_8-Q)-3HS_6,\nonumber\\
\dot S_7 &=& M_{F}S_5-(M_\times-M_\phi)S_6-3HS_7,\quad 
\dot S_8 = -\sqrt{3}M_{\phi\gamma}S_6-3HS_8\label{fin-system}.
\end{eqnarray}

\subsection{Equations of motion in absence of the Faraday effect}

The system of Eqs. \eqref{fin-system} is in the final form and we can immediately see from the equations of motion governing usual Stokes parameters, the contribution of the pseudoscalar field to the linear and circular polarization. Let us stress since now that an exact closed analytic solution for \eqref{fin-system} is not possible. However, here we consider some particular cases, by using some reasonable approximations, which allow us to find semi-analytic solutions for Eqs. \eqref{fin-system}. Indeed, the system \eqref{fin-system} can be simplified by considering the case of transverse external magnetic field, namely $\Phi=\pi/2$. This can be achieved by observing  the CMB in the direction perpendicular to the external magnetic field and for this particular configuration, the Faraday effect would be completely absent.

In the case when the Faraday effect is absent, we get the following systems of decoupled differential equations in the variable $T$
\begin{equation}\label{dec-eq}
\tilde S_1^\prime(T)=B(T)\cdot \tilde S_1(T)+(3/T)\bs I_{4\times 4}\tilde S_1(T),\qquad \tilde S_2^\prime(T)=C(T)\cdot \tilde S_2(T)+(3/T)\bs I_{5\times 5}\tilde S_2(T),
\end{equation}
where $\tilde S_1=(U, V, S_4, S_5)^\text{T}$ and $\tilde S_2=(I_\gamma, Q, S_6, S_7, S_8)^\text{T}$ are respectively two reduced (generalized) Stokes vectors\footnote{From now we omit the term generalized for the `vectors' $\tilde S_1$ and $\tilde S_2$.}. The matrices $B$ and $C$ which enter Eqs. \eqref{dec-eq} are respectively given by
\begin{eqnarray}
B(T) &=&\frac{1}{HT}\left(
\begin{matrix}
  0 & -\Delta M & M_{\phi\gamma} & 0\\
  \Delta M & 0 & 0 & M_{\phi\gamma}\\
  -M_{\phi\gamma} & 0 & 0 & -\Delta M_1\\
  0 & -M_{\phi\gamma} & \Delta M_1 & 0
 \end{matrix}\right), 
C(T)=\frac{1}{HT}\left(
\begin{matrix}
  0 & 0 & M_{\phi\gamma} & 0 & 0\\
  0 & 0 & -M_{\phi\gamma} & 0 & 0\\
  0 & M_{\phi\gamma} & 0 & -\Delta M_2 & -\sqrt{3} M_{\phi\gamma}\\
  0 & 0 & \Delta M_2 & 0 & 0\\
  0 & 0 & \sqrt{3} M_{\phi\gamma} & 0 & 0\\
   \end{matrix}\right).\nonumber
 \end{eqnarray}

\subsection{Solution of first reduced Stokes vector \texorpdfstring{$\tilde S_1$}{S1}}

Let us focus first on the solution of first reduced Stokes vector $\tilde S_1(T).$ We may note that an exact solution is not possible unless one uses some approximations that allow to find the solution by using perturbation theory, in a similar way as shown in Sec. \ref{sec:4}. Therefore, we split the matrix $B(T)$  in the following order, $B(T)=B_1(T)+B_2(T)$
\begin{eqnarray}\label{dec-matrix-2}
B_1+B_2&=&\frac{1}{HT}\left(
\begin{matrix}
  0 & 0 & M_{\phi\gamma} & 0\\
  0 & 0 & 0 & M_{\phi\gamma}\\
  -M_{\phi\gamma} & 0 & 0 & 0\\
  0 & -M_{\phi\gamma} & 0 & 0
 \end{matrix}\right)
+\frac{1}{HT}\left(
\begin{matrix}
  0 & -\Delta M & 0 & 0\\
  \Delta M & 0 & 0 & 0\\
  0 & 0 & 0 & -\Delta M_1\\
  0 & 0 & \Delta M_1 & 0
 \end{matrix}\right)
 \end{eqnarray}
where we recall that $\Delta M_1\equiv M_+-M_\phi=M_+^{\textrm{QED}}+M_+^{\textrm{CM}}+M_{\textrm{pl}}-M_\phi$. Here $M_\textrm{pl}=-\omega_\textrm{pl}^2/(2\omega)$ is the term corresponding to plasma effects which is the same for $A_+$ and $A_\times$. In order to use perturbation theory, first we must establish which part of the matrix $B$ can be treated as small perturbation.

Suppose first that matrix $B_2(T)$ can be considered as perturbation matrix, namely we can write it as the product of a small temperature independent parameter, $\epsilon$, with temperature depended functions $\tilde G(T)$ and $\tilde G_1(T)$. This situation would be true when either $|\Delta M(T)|<|\Delta M_1(T)|\ll M_{\phi\gamma}(T)$ or $|\Delta M_1(T)|<|\Delta M(T)|\ll M_{\phi\gamma}(T)$.  We will find the corresponding parameter space in Sec. \ref{sec:5}. Using the same formalism as we showed in Sec. \ref{sec:4}, we get the following solutions (to the first order in $\epsilon$) for $U$ and $V$ components of $\tilde S_1$

\begin{align}\nonumber
\left(\frac{T_i}{T}\right)^3U(T) &=\cos[F_{\phi\gamma}(T)]U_i+\left[\cos[F_{\phi\gamma}(T)]\int_{T}^{T_i} \left(G(T^\prime)\cos^2[F_{\phi\gamma}(T^\prime)]+G_1(T^\prime)\sin^2[F_{\phi\gamma}(T^\prime)]\right)dT^\prime+ \frac{1}{2}\sin[F_{\phi\gamma}(T)]\right.\nonumber\\ & \left.\times\int_T^{T_i} \Delta G(T^\prime)\sin[2F_{\phi\gamma}(T^\prime)]dT^\prime \right]V_i -\sin[F_{\phi\gamma}(T)]\,S_{4 i} -\left[\frac{1}{2}\cos[F_{\phi\gamma}(T)]\int_{T}^{T_i} \Delta G(T^\prime)\sin[2F_{\phi\gamma}(T^\prime)]dT^\prime\right.\nonumber\\ & \left.+\sin[F_{\phi\gamma}(T)]\int_{T}^{T_i} \left(G_1(T^\prime)\cos^2[F_{\phi\gamma}(T^\prime)]+G(T^\prime)\sin^2[F_{\phi\gamma}(T^\prime)]\right) dT^\prime\right]\,S_{5 i},\nonumber\\
\left(\frac{T_i}{T}\right)^3V(T) &=-\left[\cos[F_{\phi\gamma}(T)]\int_{T}^{T_i} \left(G(T^\prime)\cos^2[F_{\phi\gamma}(T^\prime)]+G_1(T^\prime)\sin^2[F_{\phi\gamma}(T^\prime)]\right)dT^\prime+ \frac{1}{2}\sin[F_{\phi\gamma}(T)]\times \right.\nonumber\\ & \left.\int_T^{T_i} \Delta G(T^\prime)\sin[2F_{\phi\gamma}(T^\prime)]dT^\prime \right]U_i+\cos[F_{\phi\gamma}(T)]V_i+\left[\frac{1}{2}\cos[F_{\phi\gamma}(T)]\int_{T}^{T_i} \Delta G(T^\prime)\sin[2F_{\phi\gamma}(T^\prime)]dT^\prime\right.\nonumber\\ & \left.+\sin[F_{\phi\gamma}(T)]\int_{T}^{T_i} \left(G(T^\prime)\sin^2[F_{\phi\gamma}(T^\prime)]+G_1(T^\prime)\cos^2[F_{\phi\gamma}(T^\prime)]\right) dT^\prime\right]\,S_{4 i}-\sin[F_{\phi\gamma}(T)]\,S_{5 i},\label{res-U-V}
\end{align}
where we have defined $F_{\phi\gamma}(T)$ and $G_1(T)$ respectively as
\begin{equation}\nonumber
F_{\phi\gamma}(T)\equiv\int_{T}^{T_i}\frac{M_{\phi\gamma}(T^\prime)}{H(T^\prime)T^\prime}\,dT^\prime,\quad G_1(T)\equiv \frac{\Delta M_1(T)}{H(T) T},
\end{equation}
and $\Delta G(T)=G(T)-G_1(T)$. Even though $\epsilon$ does not appear explicitly in \eqref{res-U-V}, it is implicitly included in $G(T)$ and $G_1(T)$. In \eqref{res-U-V} we show only the solutions for $U$ and $V$ and do not show those for the other components of $\tilde S_1$ since we are not interested in\footnote{In this paper we are only interested in usual Stokes parameters $I_\gamma, Q, U$ and $V$ since they completely describe the polarization of light. If one is also interested in intensity of pseudoscalar field $I_\phi$ which is related to $S_8$ or transition amplitudes of photons into pseudoscalar particles then are needed also expressions for the remaining Stokes parameters $S_4, S_5, S_6, S_7, S_8$.}.

So far we found the solution for $\tilde S_1$ in the case when elements of the matrix $B_1(T)$ are  much bigger in magnitude than elements of $B_2(T)$, where the last matrix has been considered as perturbation matrix. However, for some values of the parameters we have also the situation when $|\Delta M(T)|< M_{\phi\gamma}(T)\ll |\Delta M_1(T)|$. Here we are mostly interested in the case when the pseudoscalar mixing term is bigger than $|\Delta M(T)|$ because the opposite case is fulfilled  for uninteresting small values\footnote{The case $M_{\phi\gamma}(T)\ll |\Delta M(T)|$ essentially means that contribution of pseudoscalar field to the mixing is smaller than the sum of QED and CM effects. Since the last effects are very small in general, see Sec. \ref{sec:4}, the case $M_{\phi\gamma}(T)\ll |\Delta M(T)|$ is not of particular interest because it is satisfied for extremely small values of $g_{\phi\gamma}$. If indeed $g_{\phi\gamma}$ is so small, it would be very difficult to experimentally detect pseudoscalar particles, because their signal would be smaller than the QED effect even if perfect laboratory vacuum is achieved. } of $g_{\phi\gamma}$. In the case when $|\Delta M(T)|< M_{\phi\gamma}(T)\ll |\Delta M_1(T)|$ it is convenient to move the term $\Delta M(T)$ from matrix $B_2(T)$ to matrix $B_1(T)$. In this case the former matrix has non zero entries only $\Delta M_1(T)$ while the latter matrix has non zero entries $M_{\phi\gamma}(T)$ and $\Delta M(T)$. Now, the matrix $B_2(T)$ can be considered as the leading one while $B_1(T)$ can be considered as perturbation matrix. However, since $M_{\phi\gamma}(T)$ appears now in $B_1(T)$, in order to see the small effects of the pseudoscalar field, it is necessary to look for solution to the second order in $\epsilon$, namely we write  $\tilde S_1(T)= \tilde S_1^{(0)}(T)+\epsilon\tilde S_1^{(1)}(T)+\epsilon^2\tilde S_1^{(2)}(T)+ ...$ and insert it in the first equation in \eqref{dec-eq}. After collecting all terms and tedious calculations we get the following perturbative solutions for $U$ and $V$ components of $\tilde S_1$ to second order in $\epsilon$
\resizebox{1.07\linewidth}{!}{
\begin{minipage}{\linewidth}
\begin{align}
\left(\frac{T_i}{T}\right)^3U(T) &=\left(1-\int_{T}^{T_i} G(T^\prime) dT^\prime \int_{T^\prime}^{T_i}G(T^{\prime\prime}) dT^{\prime\prime}-\int_{T}^{T_i}\cos[\mathcal G_1(T^\prime)]G_{\phi\gamma}(T^\prime) dT^\prime \int_{T^\prime}^{T_i}\cos[\mathcal G_1(T^{\prime\prime})]G_{\phi\gamma}(T^{\prime\prime}) dT^{\prime\prime}\right. \nonumber\\ - & \left. \int_{T}^{T_i}\sin[\mathcal G_1(T^\prime)]G_{\phi\gamma}(T^\prime) dT^\prime \int_{T^\prime}^{T_i}\sin[\mathcal G_1(T^{\prime\prime})]G_{\phi\gamma}(T^{\prime\prime}) dT^{\prime\prime}\right)U_i +\left(\int_T^{T_i} G(T^\prime) dT^\prime+  \int_{T}^{T_i}\cos[\mathcal G_1(T^\prime)]\right.\nonumber\\ & \left.\times G_{\phi\gamma}(T^\prime) dT^\prime \int_{T^\prime}^{T_i}\sin[\mathcal G_1(T^{\prime\prime})]G_{\phi\gamma}(T^{\prime\prime}) dT^{\prime\prime} -\int_{T}^{T_i}\sin[\mathcal G_1(T^\prime)]G_{\phi\gamma}(T^\prime) dT^\prime \int_{T^\prime}^{T_i}\cos[\mathcal G_1(T^{\prime\prime})]G_{\phi\gamma}(T^{\prime\prime}) dT^{\prime\prime}\right) V_i\nonumber\\ &
+ \left(\int_T^{T_i}G(T^\prime) dT^\prime \int_{T^\prime}^{T_i}\sin[\mathcal G_1(T^{\prime\prime})]G_{\phi\gamma}(T^{\prime\prime}) dT^{\prime\prime}-\int_{T}^{T_i}\cos[\mathcal G_1(T^\prime)]G_{\phi\gamma}(T^\prime) dT^\prime\right)S_{4i} - \nonumber\\ &
\left(\int_T^{T_i}G(T^\prime) dT^\prime \int_{T^\prime}^{T_i}\cos[\mathcal G_1(T^{\prime\prime})]G_{\phi\gamma}(T^{\prime\prime}) dT^{\prime\prime}+\int_{T}^{T_i}\sin[\mathcal G_1(T^\prime)]G_{\phi\gamma}(T^\prime) dT^\prime\right)S_{5i},\nonumber\\
\left(\frac{T_i}{T}\right)^3V(T) &=- \left(\int_T^{T_i} G(T^\prime) dT^\prime+  \int_{T}^{T_i}\cos[\mathcal G_1(T^\prime)] G_{\phi\gamma}(T^\prime) dT^\prime \int_{T^\prime}^{T_i}\sin[\mathcal G_1(T^{\prime\prime})]G_{\phi\gamma}(T^{\prime\prime}) dT^{\prime\prime} -\int_{T}^{T_i}\sin[\mathcal G_1(T^\prime)]G_{\phi\gamma}(T^\prime) dT^\prime\right.\nonumber\\ & \left.\times \int_{T^\prime}^{T_i}\cos[\mathcal G_1(T^{\prime\prime})]G_{\phi\gamma}(T^{\prime\prime}) dT^{\prime\prime}\right) U_i+\left(1-\int_{T}^{T_i} G(T^\prime) dT^\prime  \int_{T^\prime}^{T_i}G(T^{\prime\prime}) dT^{\prime\prime}-\int_{T}^{T_i}\cos[\mathcal G_1(T^\prime)]G_{\phi\gamma}(T^\prime) dT^\prime \right. \nonumber\\  & \left.\times \int_{T^\prime}^{T_i}\cos[\mathcal G_1(T^{\prime\prime})]G_{\phi\gamma}(T^{\prime\prime}) dT^{\prime\prime}- \int_{T}^{T_i}\sin[\mathcal G_1(T^\prime)]G_{\phi\gamma}(T^\prime) dT^\prime \int_{T^\prime}^{T_i}\sin[\mathcal G_1(T^{\prime\prime})]G_{\phi\gamma}(T^{\prime\prime}) dT^{\prime\prime}\right)V_i\nonumber\\ & +\left(\int_T^{T_i}G(T^\prime) dT^\prime \int_{T^\prime}^{T_i}\cos[\mathcal G_1(T^{\prime\prime})]G_{\phi\gamma}(T^{\prime\prime}) dT^{\prime\prime}+\int_{T}^{T_i}\sin[\mathcal G_1(T^\prime)]G_{\phi\gamma}(T^\prime) dT^\prime\right)S_{4i} \nonumber\\ & +
\left(\int_T^{T_i}G(T^\prime) dT^\prime \int_{T^\prime}^{T_i}\sin[\mathcal G_1(T^{\prime\prime})]G_{\phi\gamma}(T^{\prime\prime}) dT^{\prime\prime}-\int_{T}^{T_i}\cos[\mathcal G_1(T^\prime)]G_{\phi\gamma}(T^\prime) dT^\prime\right)S_{5i},\label{nnr-U-V}
\end{align}
\end{minipage}}
\\
\\
where we have defined $\mathcal G_1(T)$ and $G_{\phi\gamma}(T)$ respectively as
\begin{equation}\nonumber
\mathcal G_1(T)=\int_T^{T_i} G_1(T^\prime) dT^\prime, \quad G_{\phi\gamma}(T)=M_{\phi\gamma}/(HT).
\end{equation}

\subsection{Solution of second reduced Stokes vector \texorpdfstring{$\tilde S_2$}{S2}}

Now we focus on the solution of second reduced Stokes vector $\tilde S_2$ which is the only one left. Even in this case we look for approximate solution and use perturbation theory in analogous way with the previous section. It is convenient to split the matrix $C(T)$ which enters in the second equation in \eqref{dec-eq} in the following order, $C(T)=C_1(T)+C_2(T)$
\begin{eqnarray}\label{dec-matrix-3}
C_1+C_2 &=&\frac{1}{HT}\left(
\begin{matrix}
  0 & 0 & M_{\phi\gamma} & 0 & 0\\
  0 & 0 & -M_{\phi\gamma} & 0 & 0\\
  0 & M_{\phi\gamma} & 0 & 0 & -\sqrt{3} M_{\phi\gamma}\\
  0 & 0 & 0 & 0 & 0\\
  0 & 0 & \sqrt{3} M_{\phi\gamma} & 0 & 0\\
   \end{matrix}\right)+
   \frac{1}{HT}\left(
\begin{matrix}
  0 & 0 & 0 & 0 & 0\\
  0 & 0 & 0 & 0 & 0\\
  0 & 0 & 0 & -\Delta M_2 & 0\\
  0 & 0 & \Delta M_2 & 0 & 0\\
  0 & 0 & 0 & 0 & 0\\
   \end{matrix}\right),
    \end{eqnarray}
where $\Delta M_2=M_\times -M_\phi=M_\times^\textrm{QED}+M_\times^\textrm{CM}+M_\textrm{pl}-M_\phi$. At this point we must establish which matrix in \eqref{dec-matrix-3} can be considered as perturbation matrix. This can be done by comparing the elements of $C_1(T)$ with $C_2(T)$. In the case when $|\Delta M_2(T)|\ll M_{\phi\gamma}(T)$, the matrix $C_2(T)$ can be considered as perturbation matrix and vice-versa in the case $|\Delta M_2(T)|\gg M_{\phi\gamma}(T)$.

In case when $|\Delta M_2(T)|\ll M_{\phi\gamma}(T)$, we get the following solutions\footnote{In case when the term $M_{\phi\gamma}(T)$ is much bigger than $|\Delta M_2(T)|$, it is not necessary to go beyond the first order in $\epsilon$ in perturbation theory, since the effects of the pseudoscalar field are already evident to first order in $\epsilon$.} for $I_\gamma$ and $Q$ components of $\tilde S_2$ to first order in $\epsilon$
\begin{align}
\left(\frac{T_i}{T}\right)^3I_\gamma(T) &= I_{\gamma}(T_i)+\frac{1}{2}\sin^2[F_{\phi\gamma}(T)]Q_i-\frac{1}{2}\sin[2F_{\phi\gamma}(T)]S_{6 i}+\left[\frac{1}{2}\cos[2F_{\phi\gamma}(T)]\int_T^{T_i}G_2(T^\prime)\sin[2F_{\phi\gamma}(T^\prime)]dT^\prime\right.\nonumber\\ & \left. -\frac{1}{2}\sin[2F_{\phi\gamma}(T)]\int_T^{T_i} G_2(T^\prime)\cos[2F_{\phi\gamma}(T^\prime)]\right]S_{7 i}-\frac{\sqrt{3}}{2}\sin^2[F_{\phi\gamma}(T)]S_{8 i},\nonumber\\
\left(\frac{T_i}{T}\right)^3Q(T) &= \frac{1}{4}\left(3+\cos[2F_{\phi\gamma}(T)]\right)Q_i+\frac{1}{2}\sin[2F_{\phi\gamma}(T)]S_{6 i}+\left[-\frac{1}{2}\cos[2F_{\phi\gamma}(T)]\int_T^{T_i}G_2(T^\prime)\sin[2F_{\phi\gamma}(T^\prime)]dT^\prime\right.\nonumber\\ & \left. +\frac{1}{2}\sin[2F_{\phi\gamma}(T)]\int_T^{T_i} G_2(T^\prime)\cos[2F_{\phi\gamma}(T^\prime)]dT^\prime\right]S_{7 i} +\frac{\sqrt{3}}{2}\sin^2[F_{\phi\gamma}(T)]S_{8 i},\label{res-I-Q}
\end{align}
where we defined $G_2(T)=\epsilon\tilde G_2(T)=\Delta M_2(T)/(HT)$. As in the previous section $\epsilon$ does not explicitly appear in \eqref{res-I-Q} but is implicitly included in $G_2(T)$.

The case $|\Delta M_2(T)|\gg M_{\phi\gamma}(T)$, needs a special treatment because the term corresponding to the pseudoscalar field is subdominant. In order to explore the vast region of pseudoscalar particles parameter space, it is necessary to look for solution of $\tilde S_2(T)$ up to  second order in $\epsilon$. Therefore, we expand the second reduced Stokes vector as $\tilde S_2(T)= \tilde S_2^{(0)}(T)+\epsilon\tilde S_2^{(1)}(T)+\epsilon^2 \tilde S_2^{(2)}(T)+ ...$ and insert it in the second equation in \eqref{dec-eq}. Collecting all terms, we get the following perturbative solutions for $I_\gamma$ and $Q$ components of $\tilde S_2$ to second order\footnote{As we did above, the small factor $\epsilon$ in this case in implicitly included in $M_{\phi\gamma}(T)$. } in $\epsilon$:
\begin{align}
\left(\frac{T_i}{T}\right)^3I_\gamma(T) &= I_{\gamma}(T_i)+\left(\int_T^{T_i}\cos[\mathcal{G}_2(T^\prime)]G_{\phi\gamma}(T^\prime)dT^\prime \int_{T^\prime}^{T_i}\cos[\mathcal{G}_2(T^{\prime\prime})]G_{\phi\gamma}(T^{\prime\prime})dT^{\prime\prime}+\nonumber \right.\\ & \left.\int_T^{T_i}\sin[\mathcal{G}_2(T^\prime)]G_{\phi\gamma}(T^\prime)dT^\prime \int_{T^\prime}^{T_i}\sin[\mathcal{G}_2(T^{\prime\prime})]G_{\phi\gamma}(T^{\prime\prime})dT^{\prime\prime}\right)Q_i-\left(\int_T^{T_i}\cos[\mathcal{G}_2(T^\prime)]G_{\phi\gamma}(T^\prime)dT^\prime\right) S_{6i}\nonumber\\ &-\left(\int_T^{T_i}\sin[\mathcal{G}_2(T^\prime)]G_{\phi\gamma}(T^\prime)dT^\prime\right) S_{7i}- \sqrt{3}\left(\int_T^{T_i}\cos[\mathcal{G}_2(T^\prime)]G_{\phi\gamma}(T^\prime)dT^\prime \nonumber\right.\\ & \times\left.\int_{T^\prime}^{T_i}\cos[\mathcal{G}_2(T^{\prime\prime})]G_{\phi\gamma}(T^{\prime\prime})dT^{\prime\prime}+\int_T^{T_i}\sin[\mathcal{G}_2(T^\prime)]G_{\phi\gamma}(T^\prime)dT^\prime \int_{T^\prime}^{T_i}\sin[\mathcal{G}_2(T^{\prime\prime})]G_{\phi\gamma}(T^{\prime\prime})dT^{\prime\prime}\right)S_{8i},\nonumber\\
\left(\frac{T_i}{T}\right)^3 Q(T) &= \left(1-\int_T^{T_i}\cos[\mathcal{G}_2(T^\prime)]G_{\phi\gamma}(T^\prime)dT^\prime \int_{T^\prime}^{T_i}\cos[\mathcal{G}_2(T^{\prime\prime})]G_{\phi\gamma}(T^{\prime\prime})dT^{\prime\prime}-\nonumber \right.\\ & \left. \int_T^{T_i}\sin[\mathcal{G}_2(T^\prime)]G_{\phi\gamma}(T^\prime)dT^\prime \int_{T^\prime}^{T_i}\sin[\mathcal{G}_2(T^{\prime\prime})]G_{\phi\gamma}(T^{\prime\prime})dT^{\prime\prime}\right)Q_i +\left(\int_T^{T_i}\cos[\mathcal{G}_2(T^\prime)]G_{\phi\gamma}(T^\prime)dT^\prime\right) S_{6i}\nonumber\\ & + \left(\int_T^{T_i}\sin[\mathcal{G}_2(T^\prime)]G_{\phi\gamma}(T^\prime)dT^\prime\right) S_{7i} + \sqrt{3}\left(\int_T^{T_i}\cos[\mathcal{G}_2(T^\prime)]G_{\phi\gamma}(T^\prime)dT^\prime \nonumber\right.\\ & \times\left.\int_{T^\prime}^{T_i}\cos[\mathcal{G}_2(T^{\prime\prime})]G_{\phi\gamma}(T^{\prime\prime})dT^{\prime\prime}+\int_T^{T_i}\sin[\mathcal{G}_2(T^\prime)]G_{\phi\gamma}(T^\prime)dT^\prime \int_{T^\prime}^{T_i}\sin[\mathcal{G}_2(T^{\prime\prime})]G_{\phi\gamma}(T^{\prime\prime})dT^{\prime\prime}\right)S_{8i},\label{nnr-I-Q}
\end{align}
where $\mathcal G_2(T)=\int_T^{T_i} G_2(T^\prime) dT^\prime$.

\section{Pseudoscalar particle production and generation of CMB polarization}
\label{sec:6}

In Sec. \ref{sec:5} we solved the equation of motion for the reduced Stokes vectors in case of perpendicular propagation with respect to the external magnetic field $\bs B_e$. This particular configuration, allowed us to solve the equations of motion by using perturbation theory. In this section we focus on the impact of pseudoscalar particle production in generation of CMB circular polarization.  In what follows, we concentrate mostly on generation of the CMB polarization after the decoupling epoch and estimate the degree of circular at present epoch\footnote{Our approach considered in this section is quite different from that considered in Ref. \cite{Agarwal:2008ac} where the authors consider the generation of CMB polarization for photons propagating  in magnetic field domains with fixed magnetic field amplitude and constant electron density at post decoupling time. }. In the following sections we do not consider the possibility of the rotation plane of the CMB in an external magnetic field and on the possibility of the parity in the gravitational sector. Studies in connection with the rotation angle of the polarization plane and circular polarization in an external pseudoscalar field (and not an external magnetic field) have been done in Ref. \cite{Carroll:1998zi} and studies in connection with the parity and CPT violations of the CMB have been done in Ref. \cite{Lue:1998mq}. In what follows, in order to make our treatment as simple as possible, here we concentrate for simplicity only in the case when $M_{\phi\gamma}(T)$ is subdominant with respect to the other terms $\Delta M_1(T)$ and $\Delta M_2(T)$.

\subsection{Subdominant pseudoscalar contribution: $M_{\phi\gamma}\ll |\Delta M_1|, |\Delta M_2|$}

In the case when the term $M_{\phi\gamma}(T)$ is smaller than other terms in matrices $B(T)$ and $C(T)$, which essentially corresponds to the weak mixing case, namely $M_{\phi\gamma}(T)\ll |\Delta M_{1, 2}(T)|$. In this case the expressions for the Stokes parameters $I_\gamma$ and $Q$ are given by \eqref{nnr-I-Q} while for $U$ and $V$ are given by \eqref{nnr-U-V}. Let us concentrate at the post decoupling epoch and assume that at $T=T_i$ the CMB is very weakly polarized due to Thomson scattering and consider the generation and evolution of polarization for $T\leq T_i$. In what follows, we assume that the cosmological plasma is not populated by other relic pseudoscalar particles at $T\leq T_i$, the CMB is not circularly polarized at $T=T_i$ and significant pseudoscalar particle production starts at $T=T_i$ in already existing cosmic magnetic field. In this case we have, $V_i=S_{4i}=S_{5i}=S_{6i}=S_{7i}=0$ and conservation of particle number gives $I_\gamma(T_i)=\sqrt{3}S_{8i}$. We use these values as initial conditions in expressions \eqref{nnr-U-V} and \eqref{nnr-I-Q}. In what follows we are not interested in the evolution of other Stokes parameters and will not be considered.

It is important at this stage to find the pseudoscalar parameter space that satisfy the condition of weak mixing. The cases $M_{\phi\gamma}(T)\ll |\Delta M(T)|> |\Delta M_1(T)|$ or $M_{\phi\gamma}(T)\ll |\Delta M_1(T)|> |\Delta M(T)|$ and $M_{\phi\gamma}(T)\ll |\Delta M_2(T)|$ can be solved in principle exactly, but it would be quite involved to study all possibilities of these inequality equations. However, in most practical cases it is sufficient only to calculate the leading terms in each member of the inequalities. In order to do so, let us recall that $\Delta M(T)=\Delta M_\textrm{QED}(T)+\Delta M_\textrm{CM}(T)$ and $\Delta M_1(T)=M_+^\textrm{QED}(T)+M_+^\textrm{CM}(T)+M_\textrm{pl}(T)-M_\phi(T)$. For the QED term we have essentially $|\Delta M_\textrm{QED}(T)|\sim M_+^\textrm{QED}(T)\sim M_\times^\textrm{QED}(T)$ and for the CM term, $\Delta M_\textrm{CM}(T)=-M_\times^\textrm{CM}(T)$ with $M_+^\textrm{CM}=0$ since the transverse part of external magnetic field has no $y$ component by convention. Since the plasma term is in general several orders of magnitude much bigger than QED and CM terms in $\Delta M_1$, in the parameter space that we are interested in at the post decoupling epoch, we have essentially $\Delta M_1\simeq M_\textrm{pl}-M_\phi$ where here we are also assuming that $|M_\phi|$ is bigger than QED and CM terms. Therefore, it remains to confront the term $M_\phi$ with the plasma term $M_\textrm{pl}$, where the former depends on the pseudoscalar mass $m_\phi$. Therefore we have either $|M_\phi(T)|>|M_\textrm{pl}(T)|$ or $|M_\phi(T)|<|M_\textrm{pl}(T)|$. Consequently, depending on the pseudoscalar mass, we have respectively either $|\Delta M_1(T)|\simeq |M_\phi (T)|$ or $|\Delta M_1(T)|\simeq |M_\textrm{pl}(T)|$. Being the plasma term much bigger than QED and CM terms, the previous conditions imply that in most cases we have $|\Delta M_1(T)|> |\Delta M(T)|$. Based on the same arguments one can easily show that also $\Delta M_2\simeq M_\textrm{pl}-M_\phi$.

In the weak mixing case we have $|\Delta M(T)|<|\Delta M_1(T)|$ since $|M_\times(T)|\neq |M_\phi(T)|$ and therefore it remains to find the parameter space only for $M_{\phi\gamma}(T)\ll |\Delta M_{1, 2}(T)|$ where in  most practical cases we have $|\Delta M_1|\simeq |\Delta M_2|$ for $|M_\phi| \neq |M_\textrm{pl}|$.
We find that conditions $|M_\phi|>|M_\textrm{pl}|$ and $M_{\phi\gamma}\ll |\Delta M_{1, 2}|$ are satisfied for all $T_0\leq T\leq T_i$ at the post decoupling if \footnote{It is important to stress that since we are working with perturbation theory, the conditions $M_{\phi\gamma}\ll |\Delta M_{1, 2}|$ for $|M_\phi|>|M_\textrm{pl}|$ must be satisfied in the whole interval $T_0\leq T\leq T_i$. They are respectively satisfied when their temperature dependent terms $(T_0/T)^3$ is minimum and $\sqrt{X_e(T)}(T/T_0)^{3/2}$ is maximum. On the other hand, the conditions $M_{\phi\gamma}\ll |\Delta M_{1, 2}|$ for $|M_\text{pl}|>|M_\phi |$ must be satisfied in the whole interval $T_0\leq T\leq T_i$ when the temperature dependent terms $X_{e}(T)$ and $\sqrt{X_e(T)}(T/T_0)^{3/2}$ are both minimum.}\begin{equation}\label{limit-4}
2\times 10^{-10}\quad \textrm{eV}< m_\phi,\quad g_{\phi\gamma}\ll 9.57\times 10^{15} \left(\frac{\textrm{Hz}}{\nu_0}\right)\,\left(\frac{m_{\phi}}{\textrm{eV}}\right)^{2}\,\left(\frac{\textrm{G}}{B_{e0}}\right)\,\quad \textrm{GeV}^{-1},
\end{equation}
where for $|M_\phi|>|M_\textrm{pl}|$ we have $M_{\phi\gamma}\ll |M_\phi|$. In the case when $|M_\phi|<|M_\textrm{pl}|$ we have $M_{\phi\gamma}\ll |\Delta M_{1, 2}|\simeq |M_\textrm{pl}|$ which are satisfied for all $T$ at post decoupling epoch if
\begin{equation}\label{limit-5}
m_\phi< 1.6\times 10^{-14}\,\textrm{eV},\,\quad g_{\phi\gamma}\ll 3.22\times 10^{-3}\,\bar X_e\left(\frac{\textrm{Hz}}{\nu_0}\right)\left(\frac{\textrm{G}}{B_{e0}}\right)\, \textrm{GeV}^{-1},
\end{equation}
where $\bar X_e$ is the average value of $X_e(T)$ at the post decoupling epoch. The reason of having chosen the average value will be clear below.

Now that we have established limits of validity for $I_\gamma$ and  $Q$ ($M_{\phi\gamma}(T)\ll |\Delta M_2(T)|$) and for $U$ and $V$ ($M_{\phi\gamma}(T)\ll |\Delta M_1(T)|$) in the weak mixing, we can focus on the generation of CMB circular polarization.  In the Stokes parameters $I_\gamma$ and $Q$ do appear trigonometric functions that have as argument $\mathcal G_2(T)$ while in $U$ and $V$ have as argument $\mathcal G_1(T)$. Since these functions are given respectively by the integral of $\Delta M_2(T)$ and $\Delta M_1(T)$, it may be convenient to separate if either the plasma term $M_\textrm{pl}$ or the mass term $M_\phi$ dominates in $\Delta M_{1, 2}$. As already mentioned, the QED and CM terms are much smaller than the plasma term.  Consider first the case when the plasma term $M_\textrm{pl}$ dominates in $\Delta M_{1, 2}$. Considering only the matter contribution to the Hubble parameter $H(T)$, we get
\begin{equation}\nonumber
\mathcal G_1(T)\simeq \mathcal G_2(T)=-1.56\times 10^{19} \sqrt{T_0}\left(\textrm{Hz}/\nu_0\right)\int_T^{T_i}T^{\prime -1/2}X_e(T^\prime) dT^\prime\quad (\text{K}^{-1}).
\end{equation}

We may note that $\mathcal G_{1, 2}$ is given as the integral of the inverse square root of temperature times the ionization fraction $X_e$. As already discussed in Sec. \ref{sec:4}, there is not an analytic function for $X_e$ which satisfies a complicated differential equation. In Sec. \ref{sec:4} we were able to find semi-analytic solutions for integrals involving trigonometric functions which have as argument integrals of $X_e$. For most practical cases, the argument of those trigonometric functions was much smaller than unity, but here we may note that $\mathcal G_{1, 2}$ is never less than unity for realistic values of $\nu_0$. So, in this section we cannot approximate the cosine or sine of $\mathcal G_{1, 2}$ with unity or $\mathcal G_{1, 2}$ to first order.

It is desirable to have analytic or at least semi-analytic expression for the degree of circular polarization as we did in Sec. \ref{sec:4}. Since there is no known analytic expression for $X_e$, it is convenient to replace it with its average value in $\mathcal G_{1, 2}$ at post decoupling epoch, namely $\bar X_e\simeq 0.023$. Putting $\bar X_e$ into $\mathcal G_{1, 2}$ we obtain $\mathcal G_{1, 2}(T)\simeq -1.19\times 10^{18}(\textrm{Hz}/\nu_0)(\sqrt{T_i}-\sqrt{T})$ (K$^{-1/2}$). Now with $\mathcal G_{2}(T)$ given, we can calculate the integrals which appear in $I_\gamma$ in expression \eqref{nnr-I-Q}. By integrating, we obtain 
\begin{align}\label{Integral}
\mathcal I_c=\int_{T_0}^{T_i}\cos[\mathcal{G}_{2}(T^\prime)]G_{\phi\gamma}(T^\prime)dT^\prime \int_{T^\prime}^{T_i}\cos[\mathcal{G}_{2}(T^{\prime\prime})]G_{\phi\gamma}(T^{\prime\prime})dT^{\prime\prime} &= \tilde a^{-2}\,b\left(1-\cos[105.69\,\tilde a]\right),\\ 
\mathcal I_s=\int_{T_0}^{T_i}\sin[\mathcal{G}_{2}(T^\prime)]G_{\phi\gamma}(T^\prime)dT^\prime \int_{T^\prime}^{T_i}\sin[\mathcal{G}_{2}(T^{\prime\prime})]G_{\phi\gamma}(T^{\prime\prime})dT^{\prime\prime} &= \tilde a^{-2}\,b\left(3-4\cos[52.84\,\tilde a]+\cos[105.96\,\tilde a]\right)\nonumber,
\end{align}
where we have defined $\tilde a\equiv1.19\times 10^{18}(\textrm{Hz}/\nu_0)$ and $b\equiv 6.47\times 10^{43}\left(g_{\phi\gamma}/\textrm{GeV}^{-1}\right)^2\left(B_{e0}/\textrm{G}\right)^2$. Defining $y=\mathcal I_c+\mathcal I_s$, we get the following expression for the intensity at present
\begin{equation}\label{I-weak}
(T_i/T_0)^3\,I_\gamma(T_0)=1-y+y\,Q_i,
\end{equation}
where $y \equiv 4\, \tilde a^{-2}\,b(1-\cos[52.84\,\tilde a])$ and $I_i=1$.

Let us concentrate on $V$ parameter in \eqref{nnr-U-V} and on the first term proportional to $U_i$, since other terms are absent with our choice of initial conditions. We may note the first term within parenthesis which corresponds to the QED and CM effects while other terms correspond to mixing of  pseudoscalar term with QED and CM terms. The first thing to point out, is that the term corresponding to the QED and CM effects is smaller than other terms because we are in the situation when $|\Delta M|$ is smaller than $M_{\phi\gamma}$. The second thing to note is that appear double integrals which involve sine and cosine functions in the same integral. Let $\mathcal I_{cs}$ be the double integral in the order cosine and sine functions and $\mathcal I_{sc}$ be the integral for the opposite order. In case when the plasma term $M_\textrm{pl}$ dominates $M_\phi$ in $\mathcal G_1$, is possible to find analytic expressions for $\mathcal I_{cs}$ and $\mathcal I_{sc}$ which are respectively given by
\begin{align}
\mathcal I_{cs} &=\int_{T_0}^{T_i}\cos[\mathcal{G}_{1}(T^\prime)]G_{\phi\gamma}(T^\prime)dT^\prime \int_{T^\prime}^{T_i}\sin[\mathcal{G}_{1}(T^{\prime\prime})]G_{\phi\gamma}(T^{\prime\prime})dT^{\prime\prime}= 105.69\, \tilde a^{-1}\,b\, +\nonumber \\ & \tilde a^{-2}\,b\left(\sin[105.69\,\tilde a] -4\sin[52.84\,\tilde a]\right)\nonumber,\\
\mathcal I_{sc} &=\int_{T_0}^{T_i}\sin[\mathcal{G}_{1}(T^\prime)]G_{\phi\gamma}(T^\prime)dT^\prime \int_{T^\prime}^{T_i}\cos[\mathcal{G}_{1}(T^{\prime\prime})]G_{\phi\gamma}(T^{\prime\prime})dT^{\prime\prime} =-105.69\, \tilde a^{-1}\,b+\tilde a^{-2}\,b\,\sin[105.69\,\tilde a].\nonumber
\end{align}

Now putting expressions for $\mathcal I_{cs}$ and $\mathcal I_{sc}$ in the first term in $V(T)$, we get
\begin{equation}\label{V-weak-tod}
(T_i/T_0)^3\,V(T_0)=-\left(\mathcal G(T_0)+211.38\,\tilde a^{-1}\,b -4\,\tilde a^{-2} b \sin[52.84\,\tilde a]\right)\,U_i.
\end{equation}
Using \eqref{V-weak-tod} and \eqref{I-weak} together with expression for $y$, we get the following expression for the degree of circular polarization $|V|/I_\gamma$ at $T=T_0$ to second order in perturbation theory
\begin{equation}\label{weak-circ-pol}
P_C(T_0)=\frac{|-\left(\mathcal G(T_0)+211.38\,\tilde a^{-1}\,b -4\,\tilde a^{-2} b \sin[52.84\,\tilde a]\right)\,U_i|}{1-4 b\,\tilde a^{-2}(1-\cos[52.84\tilde a])(1-Q_i)}.
\end{equation}

An important thing to note is that photon intensity must decrease and never increase in the case of pseudoscalar particle production. This happens because in our case we are not assuming photon injection in the medium by some external source that would eventually increase the photon intensity. Since $I_\gamma$ must decrease or at least remain constant, this implies that the quantity $-y+y\,Q_i$ in \eqref{I-weak} must be negative. Indeed, this is true and can be easily verified by evaluating the second and third terms in \eqref{I-weak} for a given frequency. All told, is a necessary condition but not sufficient. We must also require that $1-y+y Q_i>0$ since the intensity is a positive quantity. If we consider for example the working frequency of MIPOL experiment, $\nu_0=$ 33 GHz, for $1-y+y Q_i>0$ to be satisfied\footnote{The function $1-\cos[52.84\,\tilde a]$ in $y$ is extremely fast oscillating one and for correct evaluation for a given frequency is better to keep several digits. In this work for $\nu_0=33$ GHz we used the value of $1.82$.} we must have $|g_{\phi\gamma}|<1.66\times 10^{-15}(\textrm G/B_{e0})\quad \textrm{GeV}^{-1}$. Another consideration to be made is related with the parameter $V$ in \eqref{V-weak-tod}. As we can see in \eqref{V-weak-tod} the first term on the r. h. s. corresponds to the QED and CM effects while the second and third corresponds essentially to mixed terms. However, since perturbative expansion to second order does not reveal the asymptotic behavior of the series and because we expect that pseudoscalar contribution to $V$ to be small or $(T_i/T_0)^3 V(T_0)\lesssim U_i$, we require the additional condition that the dominant term $211.38\, \tilde a^{-1} b\lesssim 1$. In this case for $\nu_0=33$ GHz we get $|g_{\phi\gamma}|<5.13\times 10^{-20}(\textrm G/B_{e0})\quad \textrm{GeV}^{-1}$. After some algebraic operations in \eqref{weak-circ-pol} and requiring that $P_C(T_0)<7\times 10^{-5}$ (MIPOL upper limit), we get the following constraint on $g_{\phi\gamma}$ from upper limit on the degree of circular polarization 
\begin{equation}\label{mip-lim}
|g_{\phi\gamma}|<4.29\times 10^{-19}(\textrm G/B_{e0})\quad \textrm{GeV}^{-1},\quad \textrm{(MIPOL)}
\end{equation}
which is within the constraint \eqref{limit-5} and we took $U_i\simeq -Q_i$ with $Q_i\simeq 10^{-6}$. The MIPOL upper limit \eqref{mip-lim} is also satisfied if $|g_{\phi\gamma}|\lesssim 5.13\times 10^{-20}(\textrm G/B_{e0})\quad \textrm{GeV}^{-1}$ for $211.38\, \tilde a^{-1} b\lesssim 1$, which is a more conservative limit and satisfies both MIPOL upper limit and the constraint of perturbation theory.

In the domain of circular polarization, now it remains to study the last case when the term $M_\phi$ dominates the plasma term $M_\textrm{pl}$ in $\Delta M_{1, 2}$. Proceeding in the same way as above, we calculate the functions $\mathcal G_{1, 2}$ which enter the trigonometric functions in $I_\gamma$ and $V$. An important difference now is that we do not have to worry about the ionization fraction since it does not appear in $M_{\phi}$. Inserting all necessary quantities into $\mathcal G_{1, 2}$ we get
\begin{equation}\nonumber
\mathcal G_1(T)\simeq \mathcal G_2(T)=8\times 10^{47} \left(\frac{m_\phi}{\textrm{eV}}\right)^2\left(\frac{\textrm{Hz}}{\nu_0}\right)(T_i^{-5/2}-T^{-5/2})\quad (\text{K}^{5/2}).
\end{equation}
The next step is to calculate the double integrals $\mathcal I_{cs}$ and $\mathcal I_{sc}$. However, in this case there are no known analytic solutions for both type of integrals so we must evaluate them numerically together with $y$ for some specific values of the parameters.
It would be more convenient first to write $y=b\,f_1(m_\phi, \nu_0)$ and $\mathcal I_{cs}-\mathcal I_{sc}=b f_2(m_\phi, \nu_0)$ and after calculate $f_1$ and $f_2$ numerically for given values of $m_\phi$ and $\nu_0$. Let us recall that now we are in the situation in which the constraints of perturbation theory are given by \eqref{limit-4}. We may consider for example $m_\phi=10^{-8}$ eV and $\nu_0=33$ GHz which corresponds to the working frequency of MIPOL experiment. Using numerical integration we obtain $f_1=9.37\times 10^{-24}$ and $f_2=6.74\times 10^{-11}$. Second, using the relation $P_C(T_0)=|-(\mathcal G(T_0)+bf_2)U_i|/(1-y(1-Q_i))$ we get the following constraint 
\begin{equation}
|g_{\phi\gamma}|< 1.26\times 10^{-16} (\textrm{G}/B_{e0})\quad \textrm{GeV}^{-1}.
\end{equation}
If we consider for example $m_\phi=10^{-6}$ eV for the same working frequency we would obtain $f_1=9.37\times 10^{-32}$, $f_2=6.74\times 10^{-15}$ and the above limit would be two orders of magnitude weaker.

\subsection{Weak CMB polarization at decoupling}
\label{sec:6.3}

So far, in our treatment of generation of CMB polarization, we have assumed a priori that the CMB acquired a small polarization due to Thomson scattering at decoupling time, with non zero Stokes parameters $Q_i$ and $U_i$. However, even though this assumption may seems reasonable, it is not accurate because of the fact that measurements of the CMB properties are done at present epoch and not at decoupling. In fact, if cosmological magnetic fields were present at decoupling epoch, production of pseudoscalar particles would generate CMB polarization independently on Thomson scattering and also the magnetic field damping can generate temperature anisotropy and polarization \cite{Kunze:2014eka}. Consequently the linear polarization experimentally observed at present might be due to a combination of the Thomson scattering, photon-pseudoscalar particle mixing and cosmic magnetic fields\footnote{Or due to another mechanism not considered in this work.} where usually the Thomson scattering is the dominant process around the decoupling time.

Another important assumption about generation of the CMB polarization, is essentially generated at the decoupling time and it remains invariant during subsequent evolution of the universe. Even this assumption is not completely the end of the story, since for example the CMB may acquire a very small additional polarization during the reionization epoch due to Thomson scattering. So based on these two basic assumptions one would have that degree of linear polarization is generated mostly by the Thomson scattering and it remains almost invariant after the decoupling time. 

In the previous sections we found upper limits/constraints on $g_{\phi\gamma}$ from the current limit on the degree of circular polarization which is not directly generated by Thomson scattering. However, one may note that these upper limits/constraints found for $g_{\phi\gamma}$ in case of circular polarization, we have to high accuracy $P_L(T_0)\simeq P_L(T_i)$. These considerations would suggest that based on the current limit on the degree of circular polarization and consequently on derived limits/constraints on $g_{\phi\gamma}$ from it, in principle the observed CMB linear polarization could be generated by a combination of Thomson scattering and magnetic-optic effects such as photon-pseudoscalar particle mixing where the latter is the subdominant component.

In this section, we consider another possible situation, the other way round, where the contribution of the Thomson scattering to the linear polarization is supposed to be very small, so it can be completely neglected and generation of linear polarization at post decoupling time, namely at large angular scales, is mostly due to the photon-pseudoscalar particle mixing. Obviously this assumption does not mean that there is no generation of polarization due to Thomson scattering at decoupling time but simply we are neglecting this contribution to the solutions found for the Stokes parameters in the previous sections in different mixing regimes. Eventually, this approximation allows us to find weaker upper limits on the pseudoscalar particle parameter space with respect to the case where other polarization generating mechanisms would be present together with photon-pseudoscalar particle mixing at the postdecoupling time. 

Therefore based on the above assumptions, we would have that the CMB at decoupling is very weakly polarized with $Q_i\simeq 0, U_i\simeq 0, V_i\simeq 0$. Let us concentrate for the moment on the degree of linear polarization which is given by $P_L(T)=(Q^2(T)+U^2(T))^{1/2}/I_\gamma(T)$. In the weak mixing case, from \eqref{nnr-I-Q} and \eqref{nnr-U-V} and considering for example the case when $|M_\textrm{pl}|>|M_\phi|$, we get the following expression for $P_L(T)$ at present at large angular scales
\begin{equation}
P_L(T_0)=\frac{y}{1-y}\quad (\textrm{weak mixing and $|M_\textrm{pl}|>|M_\phi|$}),
\end{equation}
where we assumed the CMB unpolarized at decoupling and $U(T)=0$ for unpolarized light. Since $P_L$ depends on $y = 4\,\tilde a^{-2}\,b(1- \cos[52.84\,\tilde a])$, it explicitly depends on the photon frequency $\nu_0$. Even though one can easily calculate $P_L$ at a given frequency, is more convenient to calculate its average value on a given interval. If we consider for example $10^8$ Hz $\leq\nu_0\leq 10^{11}$ Hz, the average value of $\tilde a^{-2}\,(1-\cos[52.84\,\tilde a])\simeq 2.35\times 10^{-15}$. Considering that the degree of linear polarization of CMB at present is $P_L(T_0)\simeq 10^{-6}$, we get the following value for $\langle|g_{\phi\gamma}|\rangle\simeq 1.28\times 10^{-18} ( \textrm{G}/B_{e0})$.

In case of strong mixing, the degree of linear polarization for unpolarized CMB at decoupling is given by
\begin{equation}\label{res-lin-pol}
P_L(T_0)=\frac{(1/2) \sin^2[F_{\phi\gamma}(T_0)]}{1-(1/2)\sin^2[F_{\phi\gamma}(T_0)]},
\end{equation}
where we used expression \eqref{res-I-Q}. The condition of strong mixing, namely $M_{\phi\gamma}\gg |\Delta M_{1, 2}|$, together with $|M_\phi|>|M_\textrm{pl}|$ are satisfied for all $T_0\leq T\leq T_i$ if \footnote{The conditions $M_{\phi\gamma}\gg |\Delta M_{1, 2}|$ for $|M_\phi|>|M_\textrm{pl}|$ must be satisfied in the whole interval $T_0\leq T\leq T_i$. They are respectively satisfied when their temperature dependent terms $(T_0/T)^3$ and $\sqrt{X_e(T)}(T/T_0)^{3/2}$ are maximum. On the other hand, the conditions $M_{\phi\gamma}\gg |\Delta M_{1, 2}|$ for $|M_\text{pl}|>|M_\phi |$ must be satisfied in the whole interval $T_0\leq T\leq T_i$ when the temperature dependent terms $X_{e}(T)$ and $\sqrt{X_e(T)}(T/T_0)^{3/2}$ are respectively maximum and minimum.}
\begin{equation}\label{limit-1}
2\times 10^{-10}\quad \textrm{eV}< m_\phi,\quad g_{\phi\gamma}\gg 1.24\times 10^{25} \left(\frac{\textrm{Hz}}{\nu_0}\right)\,\left(\frac{m_{\phi}}{\textrm{eV}}\right)^{2}\,\left(\frac{\textrm{G}}{B_{e0}}\right)\,\quad \textrm{GeV}^{-1},
\end{equation}
where we used the fact that for $|M_\phi|>|M_\textrm{pl}|$, $M_{\phi\gamma}\gg |M_\phi|$. In the other case, the conditions $|M_\phi|<|M_\textrm{pl}|$ and $M_{\phi\gamma}\gg |\Delta M_{1, 2}|\simeq |M_\textrm{pl}|$ are satisfied for all  $T_0\leq T\leq T_i$ if
\begin{equation}\label{limit-2}
m_\phi< 1.6\times 10^{-14}\,\textrm{eV},\,\quad g_{\phi\gamma}\gg 3.22\times 10^{-3}\,\left(\frac{\textrm{Hz}}{\nu_0}\right)\left(\frac{\textrm{G}}{B_{e0}}\right)\, \textrm{GeV}^{-1}.
\end{equation}

The solutions of trigonometric equation \eqref{res-lin-pol} in the strong mixing together with the constrain \eqref{limit-1} (dictated by perturbation theory) are
\begin{equation}
g_{\phi\gamma}B_{e0} \simeq \left\{1.17\times 10^{-24}(2n\pi\pm 0.0014),\, 1.17\times 10^{-24}(2n\pi+3.143),\,1.17\times 10^{-24}(2n\pi+3.14)\right\}\,\text{G}\,\text{GeV}^{-1},\nonumber
\end{equation}
where $n\gg 3.14\times 10^{21}$ with $n\in \bf{Z}$ and took for simplicity $m_\phi=10^{-8}$ eV and $\nu_0=53$ GHz in \eqref{limit-1}. In the case when \eqref{limit-2} applies, the solutions of \eqref{res-lin-pol} are given by
\begin{equation}
g_{\phi\gamma}B_{e0} \simeq \left\{1.17\times 10^{-24}(2n\pi\pm 0.0014),\, 1.17\times 10^{-24}(2n\pi+3.143),\,1.17\times 10^{-24}(2n\pi+3.14)\right\}\,\text{G}\,\text{GeV}^{-1}, \nonumber
\end{equation}
where $n\gg 8.1\times 10^{9}$ with $n\in \bf{Z}$ and took for simplicity $\nu_0=53$ GHz in \eqref{limit-2}.

In the resonant case ($\Delta M_2(T)=0$), expressions for the Stokes parameters $I_\gamma$ and $Q$ are found exactly with the restrictions $g_{\phi\gamma}>0$ and the mass of the pseudoscalar particle at present must be $m_\phi(T_0)\simeq 1.6\times 10^{-14}$ eV. The expressions of Stokes parameters $I_\gamma$ and $Q$ in the resonant case coincide with those of strong mixing for $G_2(T)=0$. Consequently, expression \eqref{res-lin-pol} is valid for both resonant and strong mixing cases for unpolarized CMB at decoupling. Therefore the solutions of \eqref{res-lin-pol} in the resonant case for $g_{\phi\gamma}>0$ are 
\begin{align}\label{res-set-sol}
g_{\phi\gamma}B_{e0} &\simeq 1.17\times 10^{-24}(2n\pi- 0.0014)\,\text{G}\,\text{GeV}^{-1} \quad \text{for} \quad n\geq 1\quad  \text{or}\quad g_{\phi\gamma}B_{e0} \simeq \left\{1.17\times 10^{-24}(2n\pi+3.143)\,\text{G},\right.\nonumber \\ & \left. 1.17\times 10^{-24}(2n\pi+3.14),\, 1.17\times 10^{-24}(2n\pi+ 0.0014)\right\}\,\text{G}\,\text{GeV}^{-1}\quad \text{for}\quad n\geq 0.
\end{align}
We may note that in case when the argument of sine function in \eqref{res-lin-pol}, $F_{\phi\gamma}(T_0)$ is less than unity (or $g_{\phi\gamma} B_{e0}<1.17\times 10^{-24}\,\text{G}\,\text{GeV}^{-1}$) and because in general $P_L\ll 1$, from \eqref{res-lin-pol} one would get $\sin^2[F_{\phi\gamma}(T_0)]\simeq 2 P_L(T_0)$. Considering that $P_L(T_0)\simeq 10^{-6}$, we get in the resonant case the following frequency independent value 
\begin{equation}\label{ind-limit}
|g_{\phi\gamma}|\simeq 1.66\times 10^{-27} ( \textrm{G}/B_{e0})\, \text{GeV}^{-1} \quad \text{if}\quad g_{\phi\gamma} B_{e0}<1.17\times 10^{-24}\,\text{G}\,\text{GeV}^{-1}
\end{equation}
The solution \eqref{ind-limit} corresponds to the last set in \eqref{res-set-sol} for $n=0$. Another important thing to note in the case of linear polarization is $U(T)=0$ for initially unpolarized CMB and consequently the angle of the polarization ellipse is $\tan[2\psi(T)]=0$, which implies a horizontal linear polarization and no rotation of the polarization plane\footnote{This conclusion applies for unpolarized CMB at decoupling and for transverse magnetic field.}. In case of circular polarization, one may observe from \eqref{res-U-V} and \eqref{nnr-U-V} that $V(T)=0$ for initially unpolarized CMB, which means no generation of circular polarization for transverse magnetic field.

Mixing of CMB photons with pseudoscalar particles would also generate secondary CMB temperature anisotropy from an almost initially thermalized state\footnote{In addition to generation of CMB temperature anisotropy by photon-pseudoscalar mixing, there is also generation of temperature anisotropy by the large scale magnetic field itself. Consequently the total temperature anisotropy is given by the sum of photon-pseudoscalar mixing and magnetic field temperature anisotropy contributions. }. In order to prove this statement, consider the CMB at decoupling completely in almost thermal equilibrium and consequently almost unpolarized, $Q_i\simeq 0, U_i\simeq 0, V_i\simeq 0$. Indeed, this assumption is well motivated since $|Q_i|, |U_i|, |V_i| \ll I_i$ in both weak and strong mixing regime, see Eqs. \eqref{res-I-Q}-\eqref{nnr-I-Q}. Consider the evolution of intensity $I_\gamma$ as a function of $T$ for two specific observation directions: parallel and perpendicular to $\bs {B}_e$. In the direction parallel to $\bs B_e$, as we already have seen in Sec. \ref{sec:4}, it is induced only the Faraday effect and other magneto-optic effects are absent. As we saw there, the intensity for parallel propagation changes only due universe expansion, $(T_i/T)^3 I_\gamma(T^\parallel)=I_\gamma(T_i^\parallel)$. The intensity observed parallel to $\bs B_e$ is the same as that observed without the presence of the external field or unperturbed universe. On the other hand, the observed intensity for perpendicular propagation with respect to $\bs B_e$, for initially unpolarized CMB, is given by \eqref{res-I-Q} which in case of resonant mixing is $(T_i/T)^3I_\gamma(T^\perp)=I_\gamma(T_i^\perp)\left(1-(1/2)\sin^2[F_{\phi\gamma}(T)]\right)$. In case of thermalized CMB at decoupling we have that $I_\gamma(T_i^\perp)=I_\gamma(T_i^\parallel)=I_\gamma(T_i)$ where $I_\gamma(T_i)$ is the initial intensity at decoupling of the unperturbed black body photosphere, where for a black body $I_\gamma(T, \nu)=4\pi \nu^2\,[\exp{(2\pi\nu/T)}-1]^{-1}$. Here $T$ is the average value of the CMB over all directions.

To linear order in $\delta T$, we find the following relation from the black body intensity
\begin{equation}\label{i-t-rel}
\frac{\delta I_\gamma}{I_\gamma}=\left(\frac{x\,e^x}{e^x-1}\right)\frac{\delta T}{T_0},
\end{equation}
where we defined $x\equiv 2\pi\nu_0/T_0$ and $T_0$ is the average value of the temperature at present, averaged over all directions in the sky. In case when $x< 1$ or $\nu_0< 5.63\times 10^{10}$ Hz, namely Rayleigh-Jeans regime, we have essentially $\delta I_\gamma/I_\gamma\simeq \delta T/T_0$, while for $x>1$ (Wien regime) one must use the whole expression \eqref{i-t-rel}. We can use \eqref{i-t-rel} to find the value of $g_{\phi\gamma}$ in the resonant case\footnote{Here we consider for simplicity only the resonant case, however expression \eqref{I-aniso} is also valid in the strong mixing case. In general given the value of temperature anisotropy, the trigonometric Eq. \eqref{I-aniso} has multiple solutions in both strong and resonant mixing regimes.}. Therefore, we have
\begin{equation}\label{I-aniso}
\frac{I_\gamma(T_0^\parallel)-I_\gamma(T_0^\perp)}{I_\gamma(T_0^\parallel)}=\frac{1}{2}\sin^2[F_{\phi\gamma}(T_0)].
\end{equation}

The temperature anisotropy of the CMB depends on the angular separation between two points in the sky. Since we are comparing the intensity between parallel and perpendicular observations, this means an angular separation scale of $90^\circ$. According to WMAP9 collaboration \cite{Hinshaw:2012aka}, the temperature anisotropy at $90^\circ$ or multipole moment $ l= 2$, is $\delta T/T\simeq 3\times 10^{-5}$. From \eqref{i-t-rel} and \eqref{I-aniso} we get the following value of $g_{\phi\gamma}$
\begin{equation}\nonumber
|g_{\phi\gamma}|\simeq 9.12\times 10^{-27}\left(\frac{x\,e^x}{e^x-1}\right)^{1/2}(\text{G}/B_{e0})\quad \text{GeV}^{-1},
\end{equation}
where we considered for simplicity only the case when $F_{\phi\gamma}(T_0)<1$ and assumed that the contribution of large scale magnetic field to temperature anisotropy is subdominant to photon-pseudoscalar mixing contribution. In general \eqref{I-aniso} has multiple solutions similar to \eqref{res-set-sol} when $F(T_0)>1$.
In the Rayleigh-Jeans part of the spectrum we get $
|g_{\phi\gamma}|\simeq 9.12\times 10^{-27}(\text{G}/B_{e0})\quad \text{GeV}^{-1}$.

In the weak mixing case and for $|M_\text{pl}|>|M_\phi|$ we get
\begin{equation}\label{I-aniso-1}
\frac{I_\gamma(T_0^\parallel)-I_\gamma(T_0^\perp)}{I_\gamma(T_0^\parallel)}= y=4\,\tilde a^{-2}\,b(1-\cos[52.84\,\tilde a]),
\end{equation}
where we used the expression for $I_\gamma$ in \eqref{nnr-I-Q} with $Q_i=0$ and used the definition\footnote{$y$ should not be confused with the Compton $y$-parameter used in the CMB spectral distortion.} $y$ for $|M_\text{pl}|>|M_\phi|$. As we can see from \eqref{I-aniso-1}, $\delta I_\gamma$ is proportional to the fast varying function, $1-\cos[52.84\,\tilde a]$, which for $\tilde a=2n\pi/52.84$ is equal to zero, with $n\in \bf{Z}$, independently on the value of $g_{\phi\gamma}$. In case when $|M_\text{pl}|<|M_\phi|$, the value of $y=b f_1(m_\phi,\nu_0)$ can be calculated numerically as we did for the case of circular polarization for given values of $m_\phi$ and $\nu_0$.

\section{Discussion and conclusions}
\label{sec:7}

In this work we have studied the most important magneto-optic effects and their impact in the generation of CMB polarization. We presented a systematic study of each of them where we mostly focused on the generation of CMB circular polarization. In this work we found the equations of motion for photon and pseudoscalar fields in an external magnetic field in the WKB approximation, and then found the equations of motion for the Stokes parameters by using density matrix approach as shown in Sec. \ref{sec:3}. The resulting equations describe the mixing of different magneto-optic effects which obviously complicate the situation but on the other hand give richer scenarios.

In Sec. \ref{sec:4} we studied the vacuum polarization and CM effects separately, in order to isolate the contribution of each of them to CMB polarization. They are second order magneto-optic effects on magnetic field amplitude $B_e$ and are responsible for generation of phase shifts between the states $A_+$ and $A_\times$. These effects generate CMB elliptical polarization only in the case when the CMB is initially polarized. In this work we concentrated in the post decoupling epoch and worked under the hypothesis that the CMB acquired a small polarization at decoupling time due to Thomson scattering. We used perturbation theory and found the evolution as a function of $T$ of the Stokes parameters. We studied in particular the generation of circular polarization which is represented by the Stokes parameter $V$, in cases of observation angles $\Phi\neq\pi/2$ and $\Phi=\pi/2$.

The contribution of vacuum polarization and CM effects to $V$ depends essentially on $\Phi$, $B_{e0}$, $\nu_0$ and on the magnitude of the Stokes parameters at decoupling which, on the other hand, depend on the temperature anisotropy. In this work we assumed that $V_i=0$ at decoupling while the other parameters are non zero. The magnitude of the parameters $Q_i$ and $U_i$ obviously are smaller than temperature anisotropy and observations of CMB linear polarization give an order of magnitude of $Q_i\sim U_i\sim 10^{-6}$. 

In the case of vacuum polarization and $\Phi\neq\pi/2$, the degree of circular polarization is proportional to $Q_i$ and $U_i$, as shown in \eqref{v-today} and in most cases is the term proportional to $U_i$ which dominates. This term on the other hand is proportional to $\nu_0$ and $B_{e0}^2$. Consequently, significant generation of circular polarization would occur in the high frequency part of the CMB and for higher values of $B_{e0}$. In this work we used in our estimates a canonical value of $B_{e0}\sim $ nG but in principle higher values are possible. If for example one observes the CMB in the Wien region, say at $\nu_0\simeq 700$ GHz and the magnetic field is of the order of 100 nG, the degree of circular polarization would be of the order $P_C\sim 10^{-11}$ while for $B_{e0}\sim $ nG is four orders of magnitude smaller. 

Also for the CM effect, the degree of circular polarization is proportional to the initial values of Stokes parameters at decoupling and to $B_{e0}, \nu_0$ and $\Phi$. One distinguishing feature of the CM effect is the relation between $V_0$ and $\nu_0$ which is $V_0\propto \nu_0^{-3}$. This relation makes the CM effect quite appealing in regard to generation of circular polarization in the Rayleigh-Jeans part of the spectrum. For $\Phi\neq\pi/2$ the degree of circular polarization is given in \eqref{CM-V} where the first term is proportional to $Q_i$ and the second term is proportional to $U_i$. The term proportional to $Q_i$ shares a common feature with the vacuum polarization by the fact it gets contribution from the Faraday effect which is encoded in $\rho$. Under the approximations used in Sec. \ref{sec:4}, the term proportional to $Q_i$ is smaller than that proportional to $U_i$. The latter coincides with the solution found for $V$ in case when $\Phi=\pi/2$ for $\mathcal G(T)\ll 1$ and $F(T)<1$. This means that the contribution of the CM effect to circular polarization is bigger in the limit $\Phi\rightarrow \pi/2$. The degree of circular polarization is substantive in the frequency region $\nu_0\sim 10^8-10^9$ Hz while for higher frequencies $\nu_0\sim 10^{11}$ Hz, the contribution of CM effect to $V_0$ is subdominant to vacuum polarization. For example, if $\nu_0\sim 10^{8}$ Hz and $B_{e0}\sim $ nG, the degree of circular polarization for the CM effect would be $P_C\sim 10^{-10}$ while if $B_{e0}\sim 100$ nG, $P_C\sim 10^{-6}$.

In this work, we also studied the generation of elliptic polarization due to photon-pseudoscalar particle mixing in cosmic magnetic field, with emphasis on the degree of circular polarization. Differently from the vacuum polarization and CM effects, photon-pseudoscalar mixing has in addition two more independent parameters which are $m_\phi$ and $g_{\phi\gamma}$. We studied this mechanism in case of only transverse magnetic field and used perturbation theory to find the evolution in $T$ of the Stokes parameters. We used perturbation theory in two mixing regimes, namely weak and strong mixing and estimated the degree of circular polarization at present epoch.

Since the parameters $g_{\phi\gamma}$ and $m_\phi$ are free and in general span a wide range of values, we used the present upper limit on the degree of circular polarization in order to constrain $g_{\phi\gamma}$ and $m_\phi$. These parameters on the other hand are constrained by the mixing regimes, therefore the limits that we presented are valid in these regimes. In the strong mixing regime, in general one has to solve trigonometric equations or inequations which have as independent variable $g_{\phi\gamma} B_{e0}$. The solutions generally, depend on an integer number $n$ and consequently they are not unique. The interval of values of $g_{\phi\gamma} B_{e0}$ can in principle be narrowed by complementary constraints on $g_{\phi\gamma}$ from other methods. On the other hand, in the weak mixing case there is not such a dependence on $n$. In this case by using the upper limit on $P_C$ obtained from MIPOL experiment, we got the constraint $|g_{\phi\gamma}|<4.29\times 10^{-19}(\textrm G/B_{e0})\quad \textrm{GeV}^{-1}$ for $m_\phi<1.6\times 10^{-14}$ eV. 

Other limits have been obtained from late time (post decoupling time) generation of the degree of linear polarization and by considering the case of weakly polarized CMB at the decoupling time. The limits found in this way, obviously are weaker upper limits on the pseudoscalar particle parameter space with respect to the case where other polarization sources are present such as Thomson scattering etc. In the weak mixing case, we obtained the average value over frequency of $\langle|g_{\phi\gamma}|\rangle\sim 10^{-18} ( \textrm{G}/B_{e0})$ for $m_\phi<1.6\times 10^{-14}$ eV and $P_L(T_0)\simeq 10^{-6}$. In the strong mixing case, again one obtains values of $g_{\phi\gamma} B_{e0}$ that depends on $n$ and therefore there is no unique solution. The same thing happens even in the resonant case with the particular case that, if, $g_{\phi\gamma} B_{e0}<1.17\times 10^{-24}$, then from $P_L\simeq 10^{-6}$ we get the value  $|g_{\phi\gamma}|\simeq 1.66\times 10^{-27} ( \textrm{G}/B_{e0})$ for $m_\phi\simeq 1.6\times 10^{-14}$ eV. As in the case of vacuum polarization and CM effects, photon-pseudoscalar particle mixing generates non uniform polarization and rotation of the polarization plane across the sky. This fact can be used in order to understand if the observed linear polarization at present has a non uniform component across the sky. If this would be true, it might be due to photon-pseudoscalar particle mixing if it is the dominant mechanism of generation of linear polarization at large angular scales among the magneto-optic effects studied in this work.

From the experimental side, it turns out that among CM and QED effects, the CM effect is the most promising effect on generating circular polarization in the low frequency part of the CMB due to the dependence $V_0\propto \nu_0^{-3}$, while the vacuum polarization is the dominant one in the high frequency part due to $V_0\propto \nu_0$. The degree of circular polarization due to the CM effect in the low frequency part, in general, is bigger than that generated by vacuum polarization at high frequencies. Moreover, vacuum polarization and CM effects generate a rotation of the polarization plane of the CMB and this rotation together with the degree of circular polarization are not uniform across the sky because they depend on the observation angle $\Phi$. These facts would suggest that observation of CMB circular polarization is more likely to happen in the low frequency part of the CMB, mostly due to the CM effect and if, the observation frequency range is not a big detection issue. On the other hand, if one is interested in the measurement of the rotation angle of the polarization plane, the non uniformity of the rotation across the sky might be an issue. 

In order to detect CMB circular polarization, probably the most convenient frequency range would be for $\nu_0\sim 10^8-10^9$ Hz where the degree of circular polarization would be in the interval $P_C\simeq 10^{-13}-10^{-10}$ for $B_{e0}\simeq 1$ nG due to CM effect where the higher value of $P_C$ corresponds to the lower value of $\nu_0$. In this frequency range the contribution of vacuum polarization is completely negligible with respect to CM effect. The vacuum polarization is dominant to the CM effect in the Wien regime and for $\nu_0\sim 700$ GHz we found the interval $P_C\lesssim 10^{-15}-10^{-11}$ for the interval $B_{e0}\lesssim1-100$ nG. Assuming for the moment that frequency observation range is not an issue and it can be fixed based on experiment characteristic, one main problem on the detection of CMB circular polarization is related to the magnetic field amplitude which is poorly known. In the case of intergalactic magnetic field, usually upper limits are found by different methods and its amplitude is expected to be less than 1 nG (canonical value) up to a value of less than 100 nG. Using a canonical value of $B_{e0}\lesssim 1$ nG, one would expect that the degree of circular polarization to be  $P_C\lesssim 10^{-10}$ in the Rayleigh-Jeans regime and $P_C\lesssim 10^{-15}$ in the Wien region. These values of the degree of circular polarization corresponds essentially to the case when $\Phi=\pi/2$ and for different values of $\Phi$, $P_C$ is usually smaller and not uniform across the sky. 

The photon-pseudoscalar mixing contributes to the circular polarization as well and based on values of $g_{\phi\gamma}$ and $m_\phi$ found from the current upper limit on the circular polarization from MIPOL experiment, its contribution might be bigger than CM and vacuum polarization effects. So, let us assume for example that $B_{e0}\lesssim$ 1 nG and circular polarization would be detected with degree of circular polarization with value in the range, $10^{-10}\lesssim P_C\lesssim 10^{-6}$. This would mean there is a contribution to $P_C$ due to photon-pseudoscalar mixing which is much bigger than other magneto-optic effects or the amplitude of the magnetic field might be higher than assumed or the circular polarization is generated by another effect not considered in this work. If one would detect CMB circular polarization with average value of $P_C\lesssim 10^{-10}$, it is more likely that CM and vacuum polarization effects are the source of this polarization. 

In all studied effects, we have assumed that the large scale magnetic field was present at the decoupling epoch therefore the field has been assumed to have primordial origin and a function of spacetime coordinates, $\bs B(\bs x, t)$, namely the field is non homogeneous in space and time. In the case of vacuum polarization, the expressions for the photon polarization tensor and derived quantities such as the index of refraction have been derived under the assumption that the electromagnetic field tensor satisfies the condition $|\partial_\mu F_{\sigma\rho}|\ll m_e |F_{\sigma\rho}|$, see Ref. \cite{Dunne:2004nc} for details. This condition on the electromagnetic field tensor translates into conditions on the magnetic field amplitude $|\partial B_e^i(\bs x, t)/\partial t|\ll m_e|B_e^i(\bs x, t)|$ and $|\partial B_e^i(\bs x, t)/\partial \bs x|\ll m_e|B_e^i(\bs x, t)|$. Obviously both the last conditions on $B_e^i(\bs x, t)$ are satisfied in an expanding universe for a large class of magnetic fields where the former can be written as $H^{-1}(t)\gg 2/m_e= 7.74\times 10^{-11}$ cm for the Hubble radius as a function of time and the latter condition can be written as $l_B(t)\gg 2/m_e= 7.74\times 10^{-11}$ cm where $l_B$ is the variation scale in space of the external magnetic field. In the case of the CM effect, the elements of photon polarization tensor are usually derived for constant magnetic fields but since this effect is similar to the QED effect, namely is of the second order in $B_e$, one can extend the results for constant fields also to the case when $|\partial_\mu F_{\sigma\rho}|\ll m_e |F_{\sigma\rho}|$ in complete analogy with the vacuum polarization effect. In the case of photon-pseudoscalar mixing, the magnetic field can be either homogeneous or non homogeneous as far as the photon wavelength $\lambda\ll l_B(t)$ in the WKB approximation, namely the magnetic field is a slowly varying function in space and time with respect to the photon wavelength or frequency.

It is worth to mention also what has not been studied in this work. The first thing is related to Thomson scattering and scattering of pseudoscalar particles at post decoupling epoch, namely for $T<2970$ K and their absence in our density matrix formalism. In general, scattering is a mechanism of coherence breaking for mixing/oscillation processes which results in damping of the fields. In the density matrix formalism, the structure of the damping operator can be calculated by using field theory for scattering which is essentially the calculation of the commutator $[H_T, \rho]$ on the r. h. s. of \eqref{dens-eq} where $H_T$ includes the Hamiltonian for the Thomson scattering and that of scattering of pseudoscalar particles. However, quite often the damping term due to scattering, in case of non degenerate and non relativistic electron gas can be approximated\footnote{Similar situation occurs quite often in neutrino physics, see Ref. \cite{Dolgov:2002wy}.} by, $-i\{\Gamma, \rho\}$, where $\Gamma$ is the scattering rate matrix of photons and pseudoscalar particles which is diagonal in the basis $|A_+\rangle, |A_\times\rangle, |\phi\rangle$. Consequently, the damping term due to scattering, would have the same structure as the damping term due to Hubble friction. Therefore, the Stokes parameters would be affected by scattering but not their ratio because it cancels out exactly as the damping term due to Hubble friction. However, one must always keep in mind that this is an approximation.

The second thing is related to the case $\Phi\neq\pi/2$ for the photon-pseudoscalar particle mixing. In Sec. \ref{sec:5}, we found the equations of motion for the reduced Stokes vectors in the case of transverse external magnetic field. In this case it was possible to find two sets of decoupled differential equations for the reduced Stokes vectors and solved the equations by using perturbation theory. If the field is not transverse, namely $\Phi\neq\pi/2$, in general one has to solve simultaneously, a system of nine linear differential equations of the first order which can be problematic to solve even numerically because quite often they are stiff. We shall treat this problem in more details elsewhere but even at this stage we can outline very important conclusions about the nature of the solutions and the impact on the CMB polarization. 

The importance of solutions of the equations of motion in the case $\Phi\neq\pi/2$ (for photon-pseudoscalar mixing) relies in the fact, that being the system of equations linear, see Eqs. \eqref{fin-system}, the solutions will be proportional to initial values at a given temperature $T_i$ which does necessarily coincides with decoupling temperature. Therefore, each Stokes parameter would be proportional to $I_\gamma(T_i), Q_i, U_i, V_i, S_{4i}$ etc., and for $T<T_i$ usual Stokes parameters (those which in general interest us) would be different from zero even in case of initially unpolarized CMB at $T=T_i$. We saw similar situation in Sec. \ref{sec:6.3}, where we studied the case of unpolarized CMB at decoupling for transverse magnetic field. Consequently, the CMB would acquire polarization independently on Thomson scattering, even in case when it is initially unpolarized.

This situation would be very important in order to investigate prior decoupling CMB polarization due to photon-pseudoscalar mixing in external magnetic field. According to standard cosmology, generation of CMB polarization occurs at or very close to decoupling time due to Thomson scattering when the condition of tight coupling between photons and electro-baryon plasma breaks down. Indeed, for most models of generation of CMB polarization which include scalar perturbations, magnetic fields, gravitational waves etc., at the end is always the Thomson scattering which generates CMB polarization \cite{Zaldarriaga:2003bb}. The tight coupling condition would imply that, if there is any degree of polarization prior to decoupling, generated at temperature $T$, it would be damped very fast due to scattering of photons with electrons and baryons.  However, as we have seen and discussed in this work, photon-pseudoscalar mixing apart from generating temperature anisotropy as shown in Sec. \ref{sec:6.3} and spectral distortions of the CMB \cite{Mirizzi:2009nq}, it generates also polarization, independently on Thomson scattering. Consequently, here we advance the hypothesis that photon-pseudoscalar mixing \emph{might} generate non uniform CMB polarization across the sky, even before decoupling epoch, if the rate of photon-pseudoscalar oscillation is faster than photon scattering rate with electro-baryon plasma. Obviously, all said about this hypothesis would depend on pseudoscalar field parameters $m_\phi$ and $g_{\phi\gamma}$. The suggested hypothesis needs further attentive study and it would be too premature to conclude that it is indeed the case.

\vspace{1cm}

 {\bf{AKNOWLEDGMENTS}}:
 This work is supported by the Russian Science Foundation Grant Nr. 16-12-10037. I would like to  thank LNGS for the support received through the fellowship POR 2007-2013 `Sapere e Crescita' where part of this research was conducted.

 \appendix
 \section{Photon density operator and Stokes parameters}
 \label{app:1}

In most cases which are of interest in physics one has to deal with quantities that are proportional to the amplitude square of fields rather than the amplitude itself and that are connected with photon polarization. Such quantities are the Stokes parameters which give a complete description of  photon polarization state. Use of Stokes parameters to describe photon polarization is very convenient because allow us to deduct important information about photon polarization by using four measurable quantities associated with the photon field. The first quantity or observable expresses the intensity of the photon field while the remaining three quantities completely describe its polarization state. Stokes parameters can be applied to unpolarized, partially polarized and completely polarized light and have the mathematical convenience of not being expressed in terms of the photon amplitude, which is in general not observable. However, an observable quantity is the photon field intensity which is derived by taking the time average of the square of the amplitude.

Consider a plane wave (not necessarily monochromatic) propagating along the $z$ direction in a given cartesian coordinate system and consider the wave at $z=0$. The wave electric field vector can be decomposed along the $x$ and $y$ components as follows
$E_x(t)=E_{x0}(t)\cos[\omega t+\delta_x],\, E_y(t)=E_{y0}(t)\cos[\omega t+\delta_y],$
where $\delta_x, \delta_y$ are respectively the instantaneous wave phases for each field component, $E_{x0}$ and $E_{y0}$ are respectively the instantaneous wave amplitudes and $\omega$ is the instantaneous wave angular frequency. Here we consider the hypothesis that electric fields amplitudes 
$E_{x0}, E_{y0}$ and field phases $\delta_x, \delta_y$ slowly fluctuate in time in comparison with the rapid vibration of cosine functions. In case of nearly monochromatic wave, the Stoke's parameters are defined as follows
\begin{eqnarray}\label{Stokes-par}
I&\equiv& \langle E_{x0}^2(t)\rangle +\langle E_{y0}^2(t)\rangle,\quad Q\equiv  \langle E_{x0}^2(t)\rangle -\langle E_{y0}^2(t)\rangle,\nonumber\\
U &\equiv & \langle 2E_{x0}(t)E_{y0}(t)\cos\delta(t)\rangle,\quad V\equiv  \langle2E_{x0}(t)E_{y0}(t)\sin\delta(t)\rangle,
\end{eqnarray}
where the symbol $\langle(...)\rangle$ indicates a time average over several periods. The parameter $I$ in \eqref{Stokes-par} represents the intensity of the light, the parameter $Q$ describes the amount of linear horizontal or linear vertical polarization, $U$ describes the amount of linear polarization at an angle $\pm \pi/4$ with respect to the propagation direction and $V$ describes the amount of left or right circular polarization.

The quantities $Q$ and $U$ in general depends on the orientation of the coordinate system used for measurements but the quantities $Q^2+U^2$, $I$ and $V$ are invariant under such orientation. While the quantities $Q^2+U^2$, $I$ and $V$ are invariant under coordinate system rotation, $Q^2+U^2$ and $V$  are not necessarily invariant under simultaneous coordinate system rotation and non constant phase generation $\delta(t)$. It is straightforward to show that the orientation angle $\psi$ of the polarization ellipse can be expressed in terms of $Q$ and $U$ as
\begin{equation}\nonumber
\tan(2\psi)=U/Q,
\end{equation}
where $0\leq\psi<\pi$ physically represents the polar angle of the polarization ellipse. One can also express the total degree of polarization of the wave in terms of the Stokes parameters as follows
\begin{equation}\nonumber
P=\frac{\sqrt{Q^2+U^2+V^2}}{I},
\end{equation}
where $0\leq P\leq 1$. If $P=1$ the wave is completely polarized, if $P=0$ the wave is completely unpolarized (natural light, $Q=U=V=0$) and if $P<1$ the wave is partially polarized. The degree of linear polarization is given by $P_L=\sqrt{Q^2+U^2}/I$ and the degree of circular polarization is given by $P_C=|V|/I.$

The description of polarized light in quantum optics is different from classical optics and in general their connection is not obvious. In classical optics, polarization is described in terms of amplitudes and polarization ellipse while in quantum optics it is described in terms of the density matrix. Having defined the Stoke's parameters in \eqref{Stokes-par}, it is desirable to connect them with the polarization density matrix of a quantum optical system. As shown in Ref. \cite{Fano:1954zza}, Stokes parameters are very important tool for treating polarization problems in both quantum and classical optical systems. In order to outline their connection with the polarization density matrix, let $|A\rangle$ be an arbitrary photon state which is a linear superposition of the quantum polarization states $|A_+\rangle$ and $|A_\times\rangle$
\begin{equation}\nonumber
|A\rangle=c_1 |A_+\rangle+c_2 |A_\times\rangle,
\end{equation}
where $c_1, c_2$ are complex amplitudes. Their absolute value square represents the probability to find a photon respectively in the state $|A_+\rangle$ or $|A_\times\rangle$. In both classical and quantum optics, the polarization state of the wave is completely described in terms of complex amplitudes $c_1, c_2$ and one can define the elements of the density matrix $\rho$ as
\begin{equation}\nonumber
\rho_{ij}=c_i^*c_j, \quad (i, j=1, 2).
\end{equation}
If $F$ is any observable of the system, its expectation value on an arbitrary state is given by $\langle F\rangle=\textrm{Tr}(F_{ij}\rho_{ij})$ where summation over repeated indices is used. 

Following Ref. \cite{Fano:1954zza}, one can associate to the Stokes parameters their corresponding quantum mechanical operators as
\begin{eqnarray}
\hat I &=& |A_+\rangle\langle A_+| + |A_\times\rangle\langle A_\times|,\quad \hat U = |A_+\rangle\langle A_\times| + |A_\times\rangle\langle A_+|,\nonumber\\
\hat V &=& i\left(|A_\times\rangle\langle A_+| - |A_+\rangle\langle A_\times|\right),\quad \hat Q = |A_+\rangle\langle A_+| - |A_\times\rangle\langle A_\times|,\nonumber
\end{eqnarray}
where the expectation value of each operator is given by
\begin{eqnarray}
I &=&\langle\hat I\rangle=\textrm{Tr}(\rho_{ij}I_{ij})=\rho_{11}+\rho_{22},\quad U =\langle\hat U\rangle=\textrm{Tr}(\rho_{ij}U_{ij})=\rho_{12}+\rho_{21},\nonumber\\
V &=&\langle\hat V\rangle=\textrm{Tr}(\rho_{ij}V_{ij})=i(\rho_{12}-\rho_{21}),\quad Q =\langle\hat Q\rangle=\textrm{Tr}(\rho_{ij}Q_{ij})=\rho_{11}-\rho_{22}\label{def-stokes}.
\end{eqnarray}
One can find after some trivial algebra the polarization density matrix in the basis spanned by the photon states $|A_+\rangle, |A_\times\rangle$
\begin{equation}\label{op-density}
\rho=\frac{1}{2}\left(
\begin{matrix}
  I+Q & U-iV\\
  U+iV & I-Q\\
 \end{matrix}\right).
 \end{equation}
Expression \eqref{op-density} is an important representation of the density matrix in terms of the Stokes parameters and is very useful in many contexts. It is important to note that representation \eqref{op-density} is in the basis of the vector potential states and not in the electric field basis which is the most common used case. Consequently, the physical dimensions of the density matrix (and also Stokes parameters) in the vector potential basis are different from those in the electric field basis.


\begin{thebibliography}{99}
  
  
 \bibitem{Born65}
 A. A. Cotton and H. Mouton, ``Nouvelle propri\'{e}t\'{e} optique (bir\'{e}fringence magn\'{e}tique) de certains liquides organiques non collo\"{i}daux,'' Ct. R. hebd. Seanc Acad. Sci., Paris, {\bf 145}, 229-231 (1907)\\
A. D. Buckingam and J. A. Pople, ``A Theory of Magnetic Double Refraction,'' Proc. phys. Soc. B, {\bf 69}, 1133 (1956)\\
 M. Born,``Optik,'' Springer Berlin Heidelberg,  (1965) 602 p\\
L. D. Landau and E. M. Lifshitz, ``Electrodynamics of continuous media,'' Vol. 8, Second Edition, Pergamon (1984)
 
 
  
  
\bibitem{Heisenberg:1935qt}
  W.~Heisenberg and H.~Euler,
  ``Consequences of Dirac's theory of positrons,''
  Z.\ Phys.\  {\bf 98} (1936) 714\\
   V.~Weisskopf,
  ``The electrodynamics of the vacuum based on the quantum theory of the electron,''
  Kong.\ Dan.\ Vid.\ Sel.\ Mat.\ Fys.\ Med.\  {\bf 14N6} (1936) 1.\\  
    J.~S.~Schwinger,
  ``On gauge invariance and vacuum polarization,''
  Phys.\ Rev.\  {\bf 82} (1951) 664.\\
 R.~Karplus and M.~Neuman,
  ``The scattering of light by light,''
  Phys.\ Rev.\  {\bf 83} (1951) 776.



\bibitem{Bakalov:1994}
E.~Iacopini and E.~Zavattini,
  ``Experimental Method to Detect the Vacuum Birefringence Induced by a Magnetic Field,''
  Phys.\ Lett.\ B {\bf 85} (1979) 151.\\
D. Bakoalov {\it et al.}
Nuclear Physics B (Proceedings Supplements), {\bf{ 35}}, 180-182 (1994).\\
D. Bakoalov {\it et al.}
Quantum Semiclass. Opt., {\bf{ 10}}, Issue 1, 239-250 (1998)

\bibitem{Cameron:1993mr}
  R.~Cameron {\it et al.},
  ``Search for nearly massless, weakly coupled particles by optical techniques,''
  Phys.\ Rev.\ D {\bf 47} (1993) 3707.
  
  
\bibitem{Heinzl:2006xc}
  T.~Heinzl, B.~Liesfeld, K.~U.~Amthor, H.~Schwoerer, R.~Sauerbrey and A.~Wipf,
  ``On the observation of vacuum birefringence,''
  Opt.\ Commun.\  {\bf 267} (2006) 318  
  
  
  

\bibitem{Zavattini:2005tm}
  E.~Zavattini {\it et al.} [PVLAS Collaboration],
  ``Experimental observation of optical rotation generated in vacuum by a magnetic field,''
  Phys.\ Rev.\ Lett.\  {\bf 96} (2006) 110406
   [Phys.\ Rev.\ Lett.\  {\bf 99} (2007) 129901]



\bibitem{Zavattini:2012zs}
  G.~Zavattini, U.~Gastaldi, R.~Pengo, G.~Ruoso, F.~Della Valle and E.~Milotti,
  ``Measuring the magnetic birefringence of vacuum: the PVLAS experiment,''
  Int.\ J.\ Mod.\ Phys.\ A {\bf 27} (2012) 1260017\\
  G.~Zavattini, F.~Della Valle, A.~Ejlli and G.~Ruoso,
  ``A polarisation modulation scheme for measuring vacuum magnetic birefringence with static fields,''
  Eur.\ Phys.\ J.\ C {\bf 76} (2016) no.5,  294


\bibitem{Dafni:2006pc}
  T.~Dafni {\it et al.},
  ``First results from the CAST experiment,''
  J.\ Phys.\ Conf.\ Ser.\  {\bf 39} (2006) 117.\\
T.~Dafni {\it et al.} [IAXO and CAST Collaborations],
  ``Axion helioscopes update: the status of CAST and IAXO,''
  PoS TIPP {\bf 2014} (2014) 130



\bibitem{Asztalos:2009yp}
  S.~J.~Asztalos {\it et al.} [ADMX Collaboration],
  ``A SQUID-based microwave cavity search for dark-matter axions,''
  Phys.\ Rev.\ Lett.\  {\bf 104} (2010) 041301
  








\bibitem{Bahre:2013ywa}
  R.~BŠhre {\it et al.},
  ``Any light particle search II ÑTechnical Design Report,''
  JINST {\bf 8} (2013) T09001



  
  
  
  
  
  
\bibitem{Kim:1991zzc} 
M. J. Rees and M. Reindhardt, ``Some Remarks on Intergalactic Magnetic Fields,''
Astronomy and Astrophysics, {\bf{19}} (1972) 189\\
  K.~T.~Kim, P.~C.~Tribble and P.~P.~Kronberg,
  ``Detection of excess rotation measure due to intracluster magnetic fields in clusters of galaxies,''
  Astrophys.\ J.\  {\bf 379}, 80 (1991).\\
  J.~J.~Perry, A.~M.~Watson and P.~P.~Kronberg,
  ``Magnetic field strengths in high redshift galaxies. Can the galactic dynamo be tested?,''
  Astrophys.\ J.\  {\bf 406} (1993) 407.
  
  
\bibitem{Ade:2015cva}
  P.~A.~R.~Ade {\it et al.} [Planck Collaboration],
  ``Planck 2015 results. XIX. Constraints on primordial magnetic fields,''
  Astron.\ Astrophys.\  {\bf 594} (2016) A19
  doi:10.1051/0004-6361/201525821  
  
  
  
  
\bibitem{Neronov:1900zz}
  A.~Neronov and I.~Vovk,
  ``Evidence for strong extragalactic magnetic fields from Fermi observations of TeV blazars,''
  Science {\bf 328} (2010) 73\\ 
K.~Dolag, M.~Kachelriess, S.~Ostapchenko and R.~Tomas,
  ``Lower limit on the strength and filling factor of extragalactic magnetic fields,''
  Astrophys.\ J.\  {\bf 727} (2011) L4  
  
  
  
  
  
  \bibitem{Grasso:2000wj}
  D.~Grasso and H.~R.~Rubinstein,
  ``Magnetic fields in the early universe,''
  Phys.\ Rept.\  {\bf 348} (2001) 163\\
L.M.~Widrow,
  ``Origin of galactic and extragalactic magnetic fields,''
  Rev.\ Mod.\ Phys.\  {\bf 74}, 775 (2002) \\
M. Giovannini, ``The magnetized universe'', 
Int. J. Mod. Phys. D {\bf 13}, 391  (2004)\\
R. M. Kulsrud, E.G. Zweibel, 
``The Origin of Astrophysical Magnetic Fields'', 
Rept. Prog. Phys. {\bf 71}, 0046091 (2008) \\ 
A. Kandus, K.E. Kunze, C.G. Tsagas, 
``Primordial magnetogenesis'', 
Phys. Repts. {\bf 505}, 1 (2011) \\
R.~Durrer and A.~Neronov,
  ``Cosmological Magnetic Fields: Their Generation, Evolution and Observation,''
  Astron.\ Astrophys.\ Rev.\  {\bf 21} (2013) 62  
  
  
  
  
  
  
  
  
  
  
  
  
  
  
  
 \bibitem{Kosowsky:1996yc}
  A.~Kosowsky and A.~Loeb,
  ``Faraday rotation of microwave background polarization by a primordial magnetic field,''
  Astrophys.\ J.\  {\bf 469} (1996) 1\\ 
  L.~Campanelli, A.~D.~Dolgov, M.~Giannotti and F.~L.~Villante,
  ``Faraday rotation of the CMB polarization and primordial magnetic field properties,''
  Astrophys.\ J.\  {\bf 616} (2004) 1\\  
  C.~Scoccola, D.~Harari and S.~Mollerach,
  ``B polarization of the CMB from Faraday rotation,''
  Phys.\ Rev.\ D {\bf 70} (2004) 063003\\  
  A.~Kosowsky, T.~Kahniashvili, G.~Lavrelashvili and B.~Ratra,
  ``Faraday rotation of the Cosmic Microwave Background polarization by a stochastic magnetic field,''  
  Phys.\ Rev.\ D {\bf 71} (2005) 043006\\
  L.~Pogosian, A.~P.~S.~Yadav, Y.~F.~Ng and T.~Vachaspati,
  ``Primordial Magnetism in the CMB: Exact Treatment of Faraday Rotation and WMAP7 Bounds,''
  Phys.\ Rev.\ D {\bf 84} (2011) 043530
   [Phys.\ Rev.\ D {\bf 84} (2011) 089903]\\
   S.~De, L.~Pogosian and T.~Vachaspati,
  ``CMB Faraday rotation as seen through the Milky Way,''
  Phys.\ Rev.\ D {\bf 88} (2013) 6,  063527
  
  
  
 
 
 
 \bibitem{Bond:1987ub}
  N.~Caderni, R.~Fabbri, B.~Melchiorri, F.~Melchiorri and V.~Natale,
  ``Polarization of the microwave background radiation. I. Anisotropic cosmological expansion and evolution of the polarization states,''
  Phys.\ Rev.\ D {\bf 17} (1978) 1901.\\  
  A. G. Polnarev, Sov.Astron. {\bf{29}}, (1984) 607\\ 
  J.~R.~Bond and G.~Efstathiou,
  ``Cosmic background radiation anisotropies in universes dominated by nonbaryonic dark matter,''
  Astrophys.\ J.\  {\bf 285} (1984) L45.\\    
  J.~R.~Bond and G.~Efstathiou,
  ``The statistics of cosmic background radiation fluctuations,''
  Mon.\ Not.\ Roy.\ Astron.\ Soc.\  {\bf 226} (1987) 655.\\ 
 M.~Zaldarriaga and D.~D.~Harari,
  ``Analytic approach to the polarization of the cosmic microwave background in flat and open universes,''
  Phys.\ Rev.\ D {\bf 52} (1995) 3276\\
   U.~Seljak,
  ``Measuring polarization in cosmic microwave background,''
  Astrophys.\ J.\  {\bf 482} (1997) 6
 
\bibitem{Kosowsky:1994cy}
  A.~Kosowsky,
  ``Cosmic microwave background polarization,''
  Annals Phys.\  {\bf 246} (1996) 49
 
\bibitem{Hu:1997hv}
  W.~Hu and M.~J.~White,
  New Astron.\  {\bf 2} (1997) 323  
  
  
  
\bibitem{Kamionkowski:1996ks}
  M.~Kamionkowski, A.~Kosowsky and A.~Stebbins,
  ``Statistics of cosmic microwave background polarization,''
  Phys.\ Rev.\ D {\bf 55} (1997) 7368\\ 
 M.~Zaldarriaga and U.~Seljak,
  ``An all sky analysis of polarization in the microwave background,''
  Phys.\ Rev.\ D {\bf 55} (1997) 1830 
  
  
  
  
  
  
   
\bibitem{Crittenden:1993wm}
  R.~Crittenden, R.~L.~Davis and P.~J.~Steinhardt,
  ``Polarization of the microwave background due to primordial gravitational waves,''
  Astrophys.\ J.\  {\bf 417} (1993) L13\\
   R.~G.~Crittenden, D.~Coulson and N.~G.~Turok,
  ``Temperature - polarization correlations from tensor fluctuations,''
  Phys.\ Rev.\ D {\bf 52} (1995) 5402 \\
  K.~L.~Ng and K.~W.~Ng,
  ``Gravity wave induced polarization of the cosmic microwave background radiation,''
  Astrophys.\ J.\  {\bf 445} (1995) 521\\
  M.~Kamionkowski, A.~Kosowsky and A.~Stebbins,
  ``A Probe of primordial gravity waves and vorticity,''
  Phys.\ Rev.\ Lett.\  {\bf 78} (1997) 2058\\ 
   U.~Seljak and M.~Zaldarriaga,
  ``Signature of gravity waves in polarization of the microwave background,''
  Phys.\ Rev.\ Lett.\  {\bf 78} (1997) 2054  
  
  
  
   
 
 
 
 
 
 
  
 
 
 
\bibitem{Zaldarriaga:1998ar}
  M.~Zaldarriaga and U.~Seljak,
  ``Gravitational lensing effect on cosmic microwave background polarization,''
  Phys.\ Rev.\ D {\bf 58} (1998) 023003 
 
 
\bibitem{Mack:2001gc}
  A.~Mack, T.~Kahniashvili and A.~Kosowsky,
  ``Microwave background signatures of a primordial stochastic magnetic field,''
  Phys.\ Rev.\ D {\bf 65} (2002) 123004\\ 
  C.~Caprini, R.~Durrer and T.~Kahniashvili,
  ``The Cosmic microwave background and helical magnetic fields: The Tensor mode,''
  Phys.\ Rev.\ D {\bf 69} (2004) 063006 
 
 
 
 
\bibitem{Kovac:2002fg}
  J.~Kovac, E.~M.~Leitch, C.~Pryke, J.~E.~Carlstrom, N.~W.~Halverson and W.~L.~Holzapfel,
  ``Detection of polarization in the cosmic microwave background using DASI,''
  Nature {\bf 420} (2002) 772\\
A.~Kogut {\it et al.} [WMAP Collaboration],
  ``Wilkinson Microwave Anisotropy Probe (WMAP) first year observations: TE polarization,''
  Astrophys.\ J.\ Suppl.\  {\bf 148} (2003) 161\\
T.~E.~Montroy {\it et al.},
  ``A Measurement of the CMB  spectrum from the 2003 flight of BOOMERANG,''
  Astrophys.\ J.\  {\bf 647} (2006) 813  
 






\bibitem{negroponte:80}
J. Negroponte and J. Silk, ``Polarization of the primeval radiation in an anisotropic universe,'' Phys. Rev. Lett., {\bf{44}}, 1433 (1980)\\
M. M Basko and A. G. Polnarev, ``Polarization and anisotropy of the RELICT radiation in an anisotropic universe, '' Mon. Not. Roy. Astron. Soc., {\bf{191}}, 207 (1980)\\ 
 B. W. Tolman and R. A. Metzner, ``Large scale anisotropies and polarization of the microwave background radiation in homogeneous cosmologies,''  Proc. Roy. Soc. Lond., A {\bf{392}}, 391(1984) 
 
 
\bibitem{Alexander:2008fp}
  S.~Alexander, J.~Ochoa and A.~Kosowsky,
  ``Generation of Circular Polarization of the Cosmic Microwave Background,''
  Phys.\ Rev.\ D {\bf 79} (2009) 063524 
 

\bibitem{Bavarsad:2009hm}
  E.~Bavarsad, M.~Haghighat, Z.~Rezaei, R.~Mohammadi, I.~Motie and M.~Zarei,
  ``Generation of circular polarization of the CMB,''
  Phys.\ Rev.\ D {\bf 81} (2010) 084035


\bibitem{Giovannini:2010ar}
  M.~Giovannini,
  ``A Circular Polarimeter for the Cosmic Microwave Background,''
  JCAP {\bf 1008} (2010) 028

\bibitem{Sawyer:2012gn}
  R.~F.~Sawyer,
  ``Photon-photon interactions as a source of cosmic microwave background circular polarization,''
  Phys.\ Rev.\ D {\bf 91} (2015) no.2,  021301
  

\bibitem{De:2014qza}
  S.~De and H.~Tashiro,
  ``Circular Polarization of the CMB: A probe of the First stars,''
  Phys.\ Rev.\ D {\bf 92} (2015) no.12,  123506
  
\bibitem{Agarwal:2008ac}
  N.~Agarwal, P.~Jain, D.~W.~McKay and J.~P.~Ralston,
  ``Signatures of Pseudoscalar Photon Mixing in CMB Radiation,''
  Phys.\ Rev.\ D {\bf 78} (2008) 085028 
 
 
   
    
  
  
  
\bibitem{Mohammadi:2013dea}
  R.~Mohammadi,
  ``Evidence for cosmic neutrino background form CMB circular polarization,''
  Eur.\ Phys.\ J.\ C {\bf 74} (2014) no.10,  3102  
  
  
\bibitem{King:2016exc}
  S.~King and P.~Lubin,
  ``Circular polarization of the CMB: Foregrounds and detection prospects,''
  Phys.\ Rev.\ D {\bf 94} (2016) no.2,  023501  
  
  
  


\bibitem{smooth:83}
P. M. Lubin, P. Melese and G.F. Smooth
``Linear and circular polarization of the cosmic background radiation,''  
Astrophys. J., {\bf{273}} ,L51-L53 (1983)\\
R. B Partridge et al.,``Linear polarized fluctuations in the cosmic microwave background,''  Nature, {\bf{331}},146 (1988) 
 













\bibitem{Mainini:2013mja}
  R.~Mainini {\it et al.},
  ``An improved upper limit to the CMB circular polarization at large angular scales,''
  JCAP {\bf 1308} (2013) 033



\bibitem{Das:2004ee}
  S.~Das, P.~Jain, J.~P.~Ralston and R.~Saha,
  ``The dynamical mixing of light and pseudoscalar fields,''
  Pramana {\bf 70} (2008) 439




\bibitem{D'Olivo:2002sp}
  J.~C.~D'Olivo, J.~F.~Nieves and S.~Sahu,
  ``Field theory of the photon selfenergy in a medium with a magnetic field and the Faraday effect,''
  Phys.\ Rev.\ D {\bf 67} (2003) 025018







 \bibitem{Raffelt:1987im}
  G.~Raffelt and L.~Stodolsky,
  ``Mixing of the Photon with Low Mass Particles,''
  Phys.\ Rev.\ D {\bf 37} (1988) 1237. 
 
 
 
 
 
 
 
 
 
 
 
\bibitem{Stodolsky:1986dx}
  L.~Stodolsky,
  ``On the Treatment of Neutrino Oscillations in a Thermal Environment,''
  Phys.\ Rev.\ D {\bf 36} (1987) 2273. 
 
 
 
 
 
 
 \bibitem{Petruccione}
 H. P. Breur and F. Petruccione, ``The Theory of Open Quantum Systems'', Oxford Univ. Pr. (2002) 625 p
 
 
 
 
 
\bibitem{Kosowsky:1998mb}
  A.~Kosowsky,
  ``Introduction to microwave background polarization,''
  New Astron.\ Rev.\  {\bf 43} (1999) 157
   
 
 
 
 
\bibitem{Ade:2013zuv}
  P.~A.~R.~Ade {\it et al.}  [Planck Collaboration],
  ``Planck 2013 results. XVI. Cosmological parameters,''
  Astron.\ Astrophys.\  {\bf 571} (2014) A16 
 
  
 
 
 
 
 
 
 
\bibitem{Brezin:1971nd}
  E.~Brezin and C.~Itzykson,
  ``Polarization phenomena in vacuum nonlinear electrodynamics,''
  Phys.\ Rev.\ D {\bf 3} (1971) 618.\\
  S.~L.~Adler,
  ``Photon splitting and photon dispersion in a strong magnetic field,''
  Annals Phys.\  {\bf 67} (1971) 599.\\
  W.~Dittrich and H.~Gies,
  ``Vacuum birefringence in strong magnetic fields,''
  In *Sandansky 1998, Frontier tests of QED and physics of the vacuum* 29-43
  [hep-ph/9806417]. 
 





 
 
 
 
 
\bibitem{Weinberg:2008zzc}
 P.~J.~E.~Peebles,
  ``Recombination of the Primeval Plasma,''
  Astrophys.\ J.\  {\bf 153} (1968) 1.\\
    S.~Weinberg,
  ``Cosmology,''
  Oxford, UK: Oxford Univ. Pr. (2008) 593 p 
 
 


\bibitem{Rizzo-97}
D. M. Bishop, A. Rizzo and C. Rizzo,
``The Cotton-Mouton effect in gases: Experiment and theory",
Int. Rev. Phys. Chem. {\bf 16}, 81 (1997)



\bibitem{Carroll:1998zi}
  S.~M.~Carroll,
  ``Quintessence and the rest of the world,''
  Phys.\ Rev.\ Lett.\  {\bf 81} (1998) 3067.\\
  F.~Finelli and M.~Galaverni,
  ``Rotation of Linear Polarization Plane and Circular Polarization from Cosmological Pseudo-Scalar Fields,''
  Phys.\ Rev.\ D {\bf 79} (2009) 063002.


\bibitem{Lue:1998mq}
  A.~Lue, L.~M.~Wang and M.~Kamionkowski,
  ``Cosmological signature of new parity violating interactions,''
  Phys.\ Rev.\ Lett.\  {\bf 83} (1999) 1506.\\
  B.~Feng, M.~Li, J.~Q.~Xia, X.~Chen and X.~Zhang,
  ``Searching for CPT Violation with Cosmic Microwave Background Data from WMAP and BOOMERANG,''
  Phys.\ Rev.\ Lett.\  {\bf 96} (2006) 221302.\\
  M.~Li and X.~Zhang,
  ``Cosmological CPT violating effect on CMB polarization,''
  Phys.\ Rev.\ D {\bf 78} (2008) 103516.

\bibitem{Kunze:2014eka}
  K.~E.~Kunze and E.~Komatsu,
  ``Constraints on primordial magnetic fields from the optical depth of the cosmic microwave background,''
  JCAP {\bf 1506} (2015) no.06,  027




\bibitem{Hinshaw:2012aka}
  G.~Hinshaw {\it et al.} [WMAP Collaboration],
  ``Nine-Year Wilkinson Microwave Anisotropy Probe (WMAP) Observations: Cosmological Parameter Results,''
  Astrophys.\ J.\ Suppl.\  {\bf 208} (2013) 19





\bibitem{Dunne:2004nc}
  G.~V.~Dunne,
  ``Heisenberg-Euler effective Lagrangians: Basics and extensions,''
  In *Shifman, M. (ed.) et al.: From fields to strings, vol. 1* 445-522.






\bibitem{Dolgov:2002wy}
  A.~D.~Dolgov,
  ``Neutrinos in cosmology,''
  Phys.\ Rept.\  {\bf 370} (2002) 333


\bibitem{Zaldarriaga:2003bb}
  M.~Zaldarriaga,
  ``The polarization of the cosmic microwave background,''
  astro-ph/0305272.



\bibitem{Mirizzi:2009nq}
  A.~Mirizzi, J.~Redondo and G.~Sigl,
  ``Constraining resonant photon-axion conversions in the Early Universe,''
  JCAP {\bf 0908} (2009) 001\\
  D.~Ejlli and A.~D.~Dolgov,
  ``CMB constraints on mass and coupling constant of light pseudoscalar particles,''
  Phys.\ Rev.\ D {\bf 90} (2014) 063514



\bibitem{Fano:1954zza}
U. Fano, ``Remarks on the Classical and Quantum-Mechanical Treatment of Partial Polarization," J. Opt. Soc. Am. {\bf{39}}, 859-863 (1949)\\
D. L. Falkoff and J. E. MacDonald, ``On the Stokes Parameters for Polarized Radiation,'' J. Opt. Soc. Am. {\bf{41}}, 861(1951)\\
U.~Fano,
  ``A Stokes-Parameter Technique for the Treatment of Polarization in Quantum Mechanics,''
  Phys.\ Rev.\  {\bf 93}, 121 (1954).\\
 U.~Fano,
  ``Description of States in Quantum Mechanics by Density Matrix and Operator Techniques,''
  Rev.\ Mod.\ Phys.\  {\bf 29} (1957) 74.  
 
  






  
  
  
  
  
  
  
    












  
 \end{thebibliography}
  \end{document}